\documentclass[12pt]{article}
\usepackage{physics}
\usepackage{mathtools}
\usepackage{slashed}
\usepackage{graphicx,xcolor}
\usepackage{authblk}

\usepackage[
 	backend=bibtex,
 	style=numeric-comp,
	sorting=none
]{biblatex}
\usepackage{hyperref}
\hypersetup{
    colorlinks=true,
    linkcolor=blue,
    citecolor=violet,
    filecolor=magenta,
    urlcolor=cyan
}
\hypersetup{linktocpage}
\usepackage{orcidlink}

\usepackage[T1]{fontenc}
\usepackage{amsmath}
\usepackage{amssymb}
\usepackage{slashed}
\usepackage{color}
\usepackage{setspace}

\begin{document}

\title{\bf NLO thermal corrections to dark matter
annihilation cross sections: a novel approach}

\author[1,2]{Prabhat Butola\footnote{prabhatb@imsc.res.in,
\orcidlink{0009-0003-2824-455X}}}

\author[1,2]{D. Indumathi\footnote{indu@imsc.res.in,
\orcidlink{0000-0001-6685-4760}}}

\author[3]{Pritam Sen\footnote{pritam.sen@tifr.res.in,
\orcidlink{0000-0003-0751-5560}}}


\affil[1]{Homi Bhabha National Institute, Mumbai, India}

\affil[2]{Institute of Mathematical Sciences, Chennai, India}

\affil[3]{Department of Theoretical Physics, Tata Institute of
Fundamental Research, Mumbai, India}

\maketitle

\abstract{%
The dark matter relic density has been increasingly accurately measured by
successive generations of experiments. The Boltzmann equation determines
the yields using the dark matter annihilation cross section as one of
the inputs; the accurate computation of the latter including thermal
contributions thus assumes importance. We report here the next-to-leading
order (NLO) thermal corrections to the cross sections for (Majorana)
dark matter annihilation to standard model fermions: $\chi \chi
\to f \overline{f}$, via charged scalars. We use a novel approach,
utilising the technique of Grammer and Yennie, extended to thermal field
theories, where the cancellation of soft infra-red divergences occurs
naturally. We present the NLO thermal cross sections in full detail for
both the relativistic case as well as in the non-relativistic limit. Our
independent calculation verifies earlier results where the leading
contribution at order ${\cal{O}}(T^2)$ was shown to be proportional to
the square of the fermion mass in the non-relativistic limit, just as
at leading order. We find that the ${\cal{O}}(T^4)$ contributions have
the same dependence on the fermion mass as well.}

\vspace{0.5cm}


\section{Introduction}
\label{sec:intro}

Evidence for the presence of dark matter (DM) in our Universe arises
from several observations, primarily due to its
interaction with standard model particles via the gravitational channel
\cite{ParticleDataGroup:2022pth}. Evidence for the existence of dark
matter includes its effect in gravitational
lensing, structure formation and its presence is a necessity for
stability of rotating galaxies; various experiments have set limits
or constrained various properties of dark matter \cite{Planck:2018vyg,
Circiello:2024gpq, ManceraPina:2024ybj, Bechtol:2022koa,
Amendola:2016saw, Baudis:2015mpa} via whole-sky or galactic observations.

Cosmic microwave background radiation is sensitive to the baryonic,
dark and the total matter densities respectively, through the height of
the peaks in its power spectrum. The relic abundance of dark matter (DM)
is encoded in its relic density, currently measured to be $\Omega_c h^2
= 0.1200 \pm 0.0012$ by the PLANCK collaboration \cite{Planck:2018vyg},
where $h$ is the reduced Hubble constant, $h = H_0/100$. This is higher
than the abundance of baryonic matter and lower than the share of dark
energy in the energy density of our Universe.

Unfortunately, there is as yet no direct evidence for dark matter via,
for example, whole-sky searches of self-annihilation of dark matter
to produce gamma rays \cite{Circiello:2024gpq}, galactic surveys
\cite{ManceraPina:2024ybj}, etc. Various non-baryonic and baryonic
DM candidates have been postulated \cite{Arkani-Hamed:2006wnf,
Chen:2024njd, Arcadi:2024ukq, Bernreuther:2023kcg} such as WIMPs,
axion-like particles (ALP), neutralino dark matter, sterile neutrinos,
particles in extra dimensional theories, primordial black holes, etc.;
see Refs.~\cite{ Bauer:2017qwy, Gouttenoire:2022gwi, Roszkowski:2017nbc,
Bertone:2016nfn, Baudis:2015mpa, Garrett:2010hd} for reviews. A
large sub-set of theoretical proposals and experiments to detect
them from cosmology, colliders, etc., exist; a few recent ones can
be found in Refs.~\cite{Figueroa:2024tmn, OHare:2024nmr, Das:2024ghw,
Chakraborti:2024pdn, He:2024iju, Li:2024xlr, DelaTorreLuque:2023cef}.

One of the popular theories of dark matter consists of a SUSY-inspired model of bino-like cold dark matter
produced thermally in the Universe post-inflation. The number density
of the dark matter is then determined by Boltzmann equations
dependent on the interaction rate (annihilation/production of the dark
matter) as well as the expansion rate of the Universe. The dark matter
is in thermal, chemical and kinetic equilibrium with standard model (SM)
particles until its interaction rate falls below the Hubble expansion
rate \cite{Gondolo:1990dk}. In such a freeze-out mechanism for thermal
production of dark matter \cite{Profumo:2019ujg,Bertone:2010zza,
Profumo:2017hqp}, the DM starts decoupling from the background plasma,
when the Hubble rate of expansion of our Universe became comparable to
the annihilation rate of DM species via, for example, $\chi \chi \to f
\overline{f}$. The temperature where DM decouples from SM particles
$(f)$, depends on the mass of DM as well as the thermally averaged
annihilation cross section $\langle \sigma v_{\hbox{M\o l}} \rangle$, and we
have a range of temperature where the mass of DM can be comparable
to decoupling temperature, {\em i.e.}, $x \equiv m_\chi/T \sim
{\cal{O}}(1)$. In this range of temperature, thermal corrections can
play an important role in determining the precise relic density of dark
matter. In particular, the larger the dark matter annihilation cross
section, the later the DM goes out of equilibrium, leading to a lower
relic density, and vice versa.

A contrasting model is when the dark matter is never in equilibrium with
the SM particles; instead, the coupling of DM to SM particles is so
small that the annihilation of $\chi \chi \to f \overline{f}$
can be neglected. The amount of dark matter in the Universe then keeps
on increasing due to the reverse annihilation process $f \overline{f} \to
\chi \chi$ until the dark matter freezes-in to the present relic
density \cite{Bernal:2017kxu}. 

In any case, present-day determination of dark matter relic densities
are becoming so precise that these cross sections must necessarily
be calculated to next-to-leading order (NLO). While NLO calculations
of various relevant cross sections exist \cite{Baumgart:2022vwr,
Beneke:2020vff, Beneke:2019qaa, Klasen:2016qux, Harz:2014tma,
Laine:2022ner, Ala-Mattinen:2019mpa, Drees:2013er}, these mostly
focussed on higher order {\em quantum corrections} to the leading cross
sections in various models. It was only recently that the higher order
{\em thermal corrections} to these cross sections (or decay rates) have
been calculated \cite{Biondini:2023zcz, Binder:2021otw,
Biondini:2015gyw, Biondini:2013xua, Beneke:2014gla}. In particular,
Beneke et al.~\cite{Beneke:2014gla} computed the thermal NLO corrections
to the dark matter annihilation cross section in a bino-like model of
dark matter using thermal field theory that we describe briefly below. In
this model, which is a member of a class of such simplified models of
dark matter first discussed in Ref.~\cite{DiFranzo:2013vra}, the dark
matter candidate is an $SU(2) \times U(1)$ singlet Majorana fermion $\chi$
which interacts with SM doublet fermions, $f = (f^0,f^-)^T$, via scalar
partners $\phi = (\phi^+ , \phi^0)^T$ through a Yukawa interaction,
\begin{align}
{\cal{L}} & = -\frac{1}{4} F_{\mu\nu} F^{\mu\nu} +
\overline{{f}} \left( i \slashed{D} - m_f \right) {{f}} +
\frac{1}{2} \overline{\chi} \left( i \slashed{\partial} - m_\chi \right)
\chi \nonumber \\
  & \qquad + \left(D^\mu \phi \right)^\dagger \left(D_\mu \phi \right) -
  m_\phi^2 \phi^\dagger \phi + \left( \lambda \overline{\chi} P_L {{f}}^-
  \phi^+ + {\rm h.c.} \right)~.
\label{eq:L}
\end{align}
The thermal corrections to the annihilation of dark matter via
$\chi \chi \to f \overline{f}$ at NLO in this model were studied in
Ref.~\cite{Beneke:2014gla} where the cancellation of the infrared
divergences was explicitly demonstrated. Subsequently, an all-order
proof of the cancellation of infra-red divergences in such thermal field
theories containing scalars and fermions interacting with photons in a
heat bath at finite temperature was given in Refs.~\cite{Sen:2020oix,
Sen:2018ybx}, using a generalised Grammer and Yennie technique
\cite{Grammer:1973db,Yennie:1961ad, Indumathi:1996ec} although the finite
remainder was not calculated.  The advantage of this approach is the
automatic cancellation of the soft infra-red (IR) divergences so that we
can directly compute the finite remainder. The calculations are simpler
because they are explicitly IR finite.  Here we compute the IR-finite
contributions to the annihilation cross section at NLO, using the same
technique. We calculate both the ${\cal{O}}(T^2)$ and ${\cal{O}}(T^4)$
corrections both in the relativistic case and in the non-relativistic
limit using this novel approach to obtain an independent calculation of
the NLO thermal contributions to the annihilation cross section, and
determine the contribution in the non-relativistic limit. The latter
are important at freeze-out when $m_\chi/T \sim 20$. We show that the
${\cal{O}}(T^2)$ terms are proportional to the square of the fermion
masses, just as for the LO cross section, a result which arises due
to helicity suppression of the process due to the presence of Majorana
fermions. This result was first obtained in Ref.~\cite{Beneke:2014gla}
who subsequently calculated \cite{Beneke:2016ghp} the cross section
for annihilation of Dirac-type dark matter particles using the Operator
Product Expansion approach.

We make a careful study of the dependence of both the ${\cal{O}}(T^2)$
and ${\cal{O}}(T^4)$ corrections; the ${\cal{O}}(T^2)$ terms are indeed
proportional to $m_f^2$ in the non-relativistic limit, while there
are terms proportional to the square of the dark matter momenta $p^2$
in the general case, which of course are small in the non-relativistic
limit. We find in addition that the leading 
${\cal{O}}(T^4)$ terms are also proportional to the square of the
fermion masses in the non-relativistic limit; hence it appears that
helicity conservation determines this behaviour beyond the leading order
as well. We discuss the general behaviour of the cross section as well
as the behaviour in the non-relativistic limit; detailed expressions
for the cross section are available online \cite{online} as sets of
Mathematica Notebooks.

In the next section, we highlight some key points of thermal field
theory in the real time formalism. In Section 3, we present the
details of the Grammer and Yennie approach which allows us to easily
separate the IR-finite part of the cross section. We also discuss some
technical details about the various contributions which allow us to
simplify the calculation at NLO. In Section 4 we calculate the leading
order cross section, which is presented for completeness. In Sections 5
and 6, we calculate the NLO cross section in two different scenarios.
For both the LO and NLO calculations, we use the FeynCalc
\cite{Shtabovenko:2020gxv, Shtabovenko:2016sxi, Mertig:1990an} software
with Mathematica 13 \cite{Mathematica}. We end with Discussions in
Section 7.

\section{Thermal field theory}
\label{sec:tft}

\begin{figure}[htp]
\centering
\includegraphics[width=0.7\textwidth]{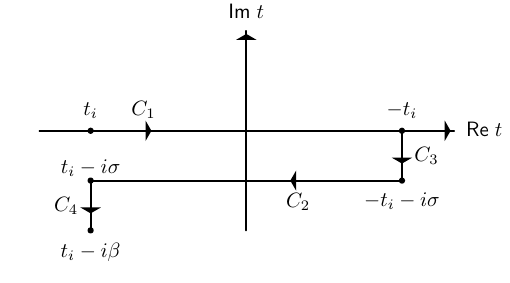}
\caption{\small \em The time path for real time formulation of thermal
field theories in the complex $t$ plane, where the $y$ axis corresponds
to $\Im t = \beta$, the inverse temperature.}
\label{fig:timepath}
\end{figure}

We briefly review the real-time formulation of thermal field theories
\cite{Kobes:1985kc, Niemi:1983ea, Rivers:1987hi, Altherr:1993tn}.
The ensemble average of an operator can be written \cite{Rivers:1987hi} as
the expectation value of time-ordered products in the thermal vacuum. The
unique feature is that the integration in the complex time plane is
defined over a special path shown in Fig.~\ref{fig:timepath}, from an
initial time, $t_i$ to a final time, $t_i - i \beta$, where $\beta$ is
the inverse temperature of the heat bath, $\beta = 1/T$;. As a result,
the thermal fields satisfy the periodic boundary conditions,
\begin{equation*}
\varphi(t_0) = \pm \varphi(t_0 - i \beta)~,
\end{equation*}
where the sign $+1$ ($-1$) corresponds to boson (fermion) fields.
The real-time axis $C_1$ allows for scattering to be defined while the
remaining portions of the contour lead to a doubling of the fields,
which can be of type-1 (the physical fields), or type-2 (the ``ghost''
fields), that live on $C_1$ and $C_2$ respectively.

Only type-1 fields can occur on external legs \cite{Altherr:1993tn} while
fields of both types can occur on internal legs, with the off-diagonal
elements of the propagator allowing for conversion of one type into
another. The Feynman rules (for the propagators and vertices) for the
Lagrangian density given in Eq.\ref{eq:L} in a heat bath at finite
temperature $T$ are listed in Appendix~\ref{sec:feyn}. We merely note
here that all propagators (which are of $2 \times 2$ matrix form), can
be written as the sum of two terms, one which is {\sf
temperature-independent} and the other which contains the explicitly thermal
dependence which we call the {\sf thermal} part.

\section{The Grammer and Yennie approach}
\label{sec:GY}

The application of the generalised Grammer and Yennie approach for the
case of the bino-like Lagrangian described in Eq.~\ref{eq:L} has been
described in detail in Refs.~\cite{Sen:2020oix, Sen:2018ybx}. Here we
summarise the main results for the sake of completeness, clarity and
use of notation.

We use a generalised version of the Grammer and Yennie technique
\cite{Grammer:1973db, Yennie:1961ad} which was used for the
zero temperature quantum field theory to separate the infrared (IR)
divergences in the theory. The technique was extended to the case of
thermal field theory, first for the case of the interaction of charged
fermions with photons in thermal QED \cite{Indumathi:1996ec}, and later
for both charged thermal scalars and a general theory of charged scalars
and fermions interacting with dark matter particles with thermal QED
corrections \cite{Sen:2020oix, Sen:2018ybx}.  In all these instances,
the all-order cancellation of infra-red (IR) divergences was shown by
separating the virtual photon propagator into two parts, the so-called $K$
and $G$ photons, which we define below, such that the IR divergence
was completely contained in the $K$ photon contribution and the $G$
photon contribution was IR finite. However, as mentioned earlier, the
IR-finite part was not explicitly calculated in Ref.~\cite{Sen:2020oix}
and is the goal of this work.

The Grammer and Yennie procedure is to insert a virtual photon with
momentum $k$ into a lower order graph, for (say) the process $A(p) B(q)
\to C(p')$, such that the momentum $q$, flowing into a given vertex $V$,
is not soft; this special vertex $V$ is labelled as a hard vertex with
respect to which the graph can be separated into a notional (initial)
$p_i$-leg and a (final) $p_f$-leg. We will identify $V$, $p_i$ and
$p_f$ for the dark matter annihilation diagrams of interest in the next
section. The $K$ and $G$ parts are subsequently identified with respect
to these definitions.  The additional inserted photon can be soft, in
which case the points of insertion will be labelled as soft vertices,
and will contribute to the IR divergence of the corresponding cross
section. The procedure starts with re-arranging, in its propagator term,
\begin{align} \nonumber -i
g^{\mu{\nu}} & \to {-i}
        \left\{\strut \left[\strut g^{\mu\nu} - b_k (p_f, p_i)\,k^\mu
        k^\nu \right] + \left[\strut b_k (p_f, p_i) k^\mu k^\nu \right]
        \right\}~, \nonumber \\
 & \equiv {-i} \left\{\left[\strut G_k^{\mu\nu}\right] +
 \left[K_k^{\mu\nu}\right] \right\}~,
\label{eq:KG}
\end{align}
where
\begin{align}
b_k(p_f,p_i) = \frac{1}{2} \left[ \frac{(2p_f-k) \cdot
(2p_i-k)}{((p_f-k)^2-m^2)((p_i-k)^2-m^2)} + (k \leftrightarrow -k)
\right]~,
\label{eq:bk}
\end{align}
is defined symmetrically in $k\to -k$ for the thermal case (in contrast
to the original definition in Ref.~\cite{Grammer:1973db}), and is a
function of $k$ as well as the momenta, $p_f$, $p_i$.
It can then be shown that the $K$ photon insertions contain the IR
divergent pieces while the $G$ photon contribution is IR finite.

\subsection{The $K$ photon insertion and the IR divergent piece}

\begin{figure}[htp]
\begin{center}
\includegraphics[width=0.35\textwidth]{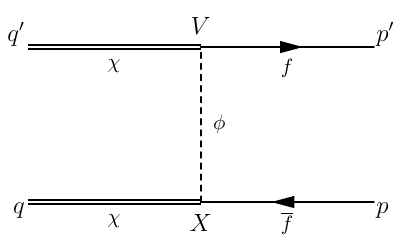}
\hspace{1cm}
\includegraphics[width=0.35\textwidth]{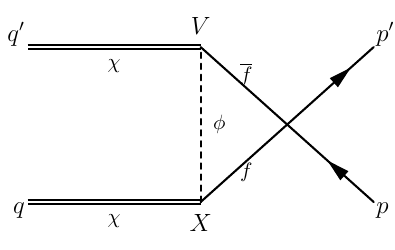}
\caption{\small \em The $t$-channel and $u$-channel dark matter
annihilation processes at leading order (LO).}
\label{fig:lo}
\end{center}
\end{figure}

We briefly describe, for the sake of completeness, the separation and
cancellation of IR divergences in the theory. Details are available in
Refs.~\cite{Sen:2020oix, Sen:2018ybx}. Let us start with the
example of insertion of an additional virtual photon in the lowest-order
diagrams shown in Fig.~\ref{fig:lo}. The key step is to specifically
define a ``hard vertex'' $V$ with respect to which we can identify the
momenta $p_f$ and $p_i$. We (arbitrarily) define this special vertex $V$
to be the one where the momentum $q'$ (of $\chi$) enters. Then, a higher
order diagram can be obtained by inserting a virtual photon anywhere in
either of the LO diagrams shown in Fig.~\ref{fig:lo}. For every such
insertion, $p_f$ ($p_i$) will correspond to the momentum $p'$ or $p$
depending on whether the final (initial) vertex of the virtual photon
was inserted on the $p'$ or $p$ leg.  Then for example, a virtual photon
inserted with one vertex on the final fermion leg and one on the scalar
leg (as in Diagram (1) of Fig.~\ref{fig:nlo}), will have a virtual photon
propagator involving the term $b_k(p',p)$ since the $p_f$ line corresponds
to the $p'$ leg, while the $p_i$ line corresponds to the scalar leg
from $V$ to $X$ together with the anti-fermion leg with momentum $p$.
Thus the corresponding $K$ and $G$ photon insertions can be defined
using Eq.~\ref{eq:KG}. For the $u$-channel diagram, such an insertion
will involve $b_k(p',p')$; see Ref.~\cite{Sen:2020oix} for more details.

We now demonstrate the factorisation of the IR divergences in the
$K$-photon insertions. We take the example of the case when the inserted
photon has both its vertices, $\mu, \nu$, on the $p'$ fermion line. The
relevant part of the matrix element can be expressed as (note that there
is an integration measure over the photon momentum, $k$)
\begin{align}
M_{NLO}^{p'p', K{\rm photon}} & \propto
b_k(p',p') k^\mu \, k^\nu\, \left[\overline{u}({p'},{m_f}) \gamma_\mu
\, {\cal{S}}_{\rm fermion}^{t_\mu,t_\nu}(p'+k, m_f) \gamma_\nu \,
{\cal{S}}_{\rm fermion}^{t_\nu,t_V} (p', m_f)
P_R \, u({q'},{m_\chi}) \right] \cdots ~, \nonumber \\
& = b_k(p',p') \, \left[\overline{u}_{p'} \, \slashed{k}
\, {\cal{S}}^{\mu,\nu}_{p'+k} \, \slashed{k} \, {\cal{S}}^{\nu,V}_{p'}
P_R \, u_{q'} \right] \cdots ~,
\label{eq:Kpppp}
\end{align}
where the ellipsis refer to terms independent of $k$ and we have
lightened the notation for convenience in the second line, with the
superscripts on ${\cal{S}}$ referring to the thermal type.  The term
in the square brackets can be simplified, using the identities given in
Appendix~\ref{app:fidentities} as
\begin{align}
\left[ \hspace{0.5cm} \right] & = \left[\overline{u}_{p'} \, \slashed{k}
\, \left\{{\cal{S}}^{\mu,\nu}_{p'+k} \, \slashed{k}
{\cal{S}}^{\nu,V}_{p'} \right\} \,
P_R \, u_{q'} \right]~, \nonumber \\
& = (-1)^{t_\nu+1} \, \left[\overline{u}_{p'} \, \slashed{k}
\, \left\{ {\cal{S}}_{p'}^{\mu,V} \, \delta_{t_\nu,t_V} -
{\cal{S}}_{p'+k}^{\mu,V} \, \delta_{t_\nu,t_\mu} \right\}
P_R \, u_{q'} \right]~, \nonumber \\
& = (-1)^{t_\nu+1} \, \left[\overline{u}_{p'}
\, \left\{ 0 -\left[(-1)^{t_\mu+1} \, \delta_{t_\nu,t_\mu} \right] \right\}
P_R \, u_{q'} \right]~,
\label{eq:Kdiff}
\end{align}
where the first term vanishes since the integrand is odd in $k \to -k$
while both $b_k$ and the measure are even under this exchange. The crucial
step is seen in the second line of Eq.~\ref{eq:Kdiff}, where the Feynman
identities (see Eq.~\ref{eq:endf}) are used to reduce the portions with
$\slashed{k}$ insertions to a {\em difference} of two terms. Hence the
matrix element factorises into two terms, one that is proportional to
the lower order matrix element, {\em viz.},
\begin{align}
M_{NLO}^{p'p', K{\rm photon}} & \propto b_k(p',p') \, M_{LO}~,
\label{eq:pppp}
\end{align}
and the other containing the IR divergence. The contribution to the cross
section from virtual photon insertions at a given order is obtained by
adding all the matrix elements upto that order and squaring the sum to obtain
$\vert {\cal{M}} \vert^2$, integrating over the appropriate phase
space, and dividing by the initial flux, as usual; more details are
given in Section~\ref{sec:cr_nlo_nondyn}.

\subsubsection{Insertion into a general $n^{\rm th}$ order graph}

Consider an $n^{\rm th}$ order graph with $n$ virtual/real photon
vertices; for specificity, we consider higher order corrections to
the $t$-channel diagram. When an additional virtual $K$-photon is
inserted into this graph, there are several possible locations where the
additional photon vertices can be inserted. Adding all the contributions
gives differences of two terms as seen above, with, more importantly,
sets of terms cancelling against each other until only one term, that
is proportional to the lower order matrix element, is left.

A similar result is obtained when the photon vertices are inserted so
that one vertex $\mu$ is on the $p'$ leg and the other $\nu$ vertex is
on the $p$ leg; that is, either on the scalar line or the anti-fermion
line. Here again, it turns out that the $K$ photon contribution is
proportional\footnote{Technically, a virtual photon insertion leads to
the number of vertices increasing by 2; we have used the index $(n+1)$
to indicate that it is an $n^{\rm th}$ order graph with an additional
virtual photon.} to the lower order matrix element:
\begin{align}
M_{n+1}^{p'p, K{\rm photon}} & \propto b_k(p',p) \, M_{n}~,
\label{eq:ppp}
\end{align}
In fact, such an insertion is the sum of contributions when the
second $\nu$ vertex of the virtual photon is inserted on the scalar,
and when it is inserted on the anti-fermion line. These contributions
separately cancel among themselves, leaving one term in the former and
two in the latter. These cancel across the vertex $V$, leaving behind
a contribution that is again proportional to the lower order graph, as
seen in Eq.~\ref{eq:ppp} above. In short, it is found \cite{Sen:2020oix}
that the total matrix element due to insertion of the virtual $K$ photon
into an $n^{\rm th}$ order diagram is given by
\begin{align} \nonumber
{\cal{M}}_{n+1}^{K{\rm photon, tot}} = & ~\left[ \frac{ie^2}{2} \int
	\frac{{\rm d}^4 k}{(2\pi)^4} \, \left\{
	\rule{0pt}{12pt} \delta_{t_\mu,t_1} \,
	\delta_{t_\nu,t_1} \, D^{t_\mu,t_\nu} (k) \,
	\left[\strut b_k(p',p') + b_k(p,p) \right] \right. \right. \nonumber \\
	 & ~~~~+\left. \left. \delta_{t_\mu,t_V} \, \delta_{t_\nu,t_V} \,
	 D^{t_\mu,t_\nu} (k) \,
	\left[\strut -2b_k(p',p) \right] \rule{0pt}{12pt} \right\}
	\rule{0pt}{16pt} \right]
	 {\cal{M}}_{n}~, \nonumber \\
	 \equiv & ~\left[B \right] {\cal{M}}_{n}~,
\label{eq:K}
\end{align}
where the prefactor containing the IR divergence can be expressed as,
\begin{align} \nonumber
B & = \frac{ie^2}{2} \int
	\frac{{\rm d}^4 k}{(2\pi)^4} \, D^{11} (k) \,
	\left[\strut b_k(p',p') - 2 b_k(p',p) + b_k(p,p) \right]~, 
	 \nonumber \\
  & \equiv \frac{ie^2}{2} \int
	\frac{{\rm d}^4 k}{(2\pi)^4} \,
	D^{11} (k) \, \left[\strut J^2(k) \right]~,
\label{eq:B}
\end{align}
since the thermal types of the hard/external vertices must be type-1. We
see that each term is proportional to the (11) component of the photon
contribution and this is crucial for achieving the cancellation between
virtual and real photon insertions.

When the contributions are summed over all orders in perturbation
theory and squared, the relevant IR divergent factor $(B+B^*)$
exponentiates and cancels a similar IR divergent factor arising
from real soft photon insertions; for details please refer to
Ref.~\cite{Sen:2020oix}.

\subsubsection{Insertion of the photon into a thermal fermion line}
\label{sec:thermalfermion}

\begin{figure}[htp]
\begin{center}
\includegraphics[width=0.6\textwidth]{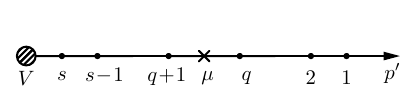}
\caption{\small \em Sample insertion of vertex $\mu$ of virtual photon
between vertices $\mu_q$ and $\mu_{q+1}$ on the $p'$ fermion line. The
labels have been simplified ($\mu_i \to i$) for the sake of clarity. Only
a portion of the diagram containing the $p'$ leg to the right of vertex
$V$ has been shown here.}
\label{fig:finite}
\end{center}
\end{figure}

We already know that the presence of the bosonic number operator,
$n_B(\vert k^0 \vert)$, makes the IR contribution to the cross section
potentially IR divergent while the nature of the fermionic number
operator, $n_F(\vert p^0 \vert)$, results in a finite IR contribution to
the cross section. In particular, we shall see below that the $K$-photon
contribution when the additional photon is inserted into a {\sf thermal}
fermion (that is, when the second, temperature-dependent term in the
fermionic propagator contributes; see Eq.~\ref{eq:fprop}), is zero. We
will use this result later in the computation of the NLO cross section
of interest. In order to show this, we begin with the insertion of one
of the vertices of the additional $K$ photon on any thermal fermion line
at the vertex $\mu$, lying between vertices $\mu_{q+1}$ and $\mu_q$;
see Fig.~\ref{fig:finite}.

There are now two fermion propagators, one between vertices $\mu_q$ and $\mu$
and the other between vertices $\mu$ and $\mu_{q+1}$. The relevant part of
the matrix element is given by (where we have chosen to insert into the
$p'$ fermion line for specificity),
\begin{align} M_{NLO}^{{\rm thermal~f}, K\gamma} &
\propto b(p_f, p_i) k^\mu k^\nu \left[ \overline{u}_{p'} \gamma_{\mu_1}
\cdots \gamma_{\mu_q} \, {\cal{S}}^{q,\mu}_{p'+\sum_q} \gamma_\mu \,
{\cal{S}}^{\mu,q+1}_{p'+\sum_q+k} \, \gamma_{\mu_{q+1}} \right] \cdots~,
\nonumber \\
 & = b(p_f, p_i) k^\nu \left[ \overline{u}_{p'} \gamma_{\mu_1} \cdots
\gamma_{\mu_q} \, \left\{ {\cal{S}}^{q,\mu}_{p'+\sum_q} \slashed{k} \,
{\cal{S}}^{\mu,q+1}_{p'+\sum_q+k} \right\} \gamma_{\mu_{q+1}} \right]
\cdots~,
\end{align}
where the momentum flowing in the fermion line between vertices $q$
and $(q+1)$ into which the additional photon was inserted is $p' +
l_1 + \cdots + l_q \equiv p' + \Sigma_q$. The term in curly braces
can be simplified using the property of the fermion propagator shown
in Eq.~\ref{eq:fprop} where $\overline{S}$ is also defined:
\begin{align}
\left\{ \cdots \right\}
  & = \left[(\slashed{p'} + \slashed{\Sigma}_q + m_f) \slashed{k}
(\slashed{p'} + \slashed{\Sigma}_q + \slashed{k} + m_f) \right]
\overline{S}^{q,\mu}_{p'+\sum_q} \overline{S}^{\mu,q+1}_{p'+\sum_q+k}~,
\nonumber \\
  & = \left[ (2(p' + \Sigma_q) \cdot k + k^2) \,(\slashed{p'}
+ \slashed{\Sigma}_q + m_f) - ((p'+\Sigma_q)^2- m_f^2) \slashed{k}
\right]~, \nonumber \\
  & = 0~ \hbox{ for thermal fermions}.
\label{eq:thermalfermionK}
\end{align}
In particular, if both the fermion propagators on either side of the
$K$-photon insertion at the vertex $\mu$ are {\sf thermal}, then both
terms in Eq.~\ref{eq:thermalfermionK} vanish due to the delta function
terms $\delta ((p'+\Sigma_q)^2- m_f^2)$ and $\delta ((p'+\Sigma_q+k)^2-
m_f^2)$ in the thermal part of the respective propagators. We will use
this result in the next section when we compute the NLO cross section.

\subsubsection{The thermal real photon corrections}
\label{sec:realphoton}

In order to show the IR divergence cancellation between virtual and
real photon contributions, a similar treatment of the real photon
corrections to the process is done, by separating the polarisation
sums for the insertion of real photons into so-called $\widetilde{K}$
and $\widetilde{G}$ contributions:
\begin{equation}
\begin{array}{rcl}
\displaystyle \sum\limits_{\rm pol} \epsilon^{*,\mu} (k)\,\epsilon^\nu (k)
& = & \displaystyle -
	g^{\mu\nu}~, \\
 & = & \displaystyle -\left\{\strut \left[\strut g^{\mu\nu} -
	\tilde{b}_k(p_f,p_i) k^\mu k^\nu \right] +
	\left[\strut \tilde{b}_k(p_f,p_i) k^\mu k^\nu \right] \right\}~,
	 \\[2ex]
 & \equiv & \displaystyle -\left\{\left[ \widetilde{G}_k^{\mu\nu}\right] + \left[
 \widetilde{K}_k^{\mu\nu}\right]\right\}~.
\end{array}
\label{eq:KGtilde}
\end{equation}
Since $k^2=0$ for real photons, we define,
\begin{equation}
\tilde{b}_k(p_f, p_i) = b_k(p_f, p_i)\Big|_{k^2 = 0} =
	\frac{p_f \!\cdot \!p_i}{ k\!\cdot \!p_f\; k\!\cdot \!p_i}~,
\end{equation}
where $p_i$ ($p_f$) is the momentum $p'$ or $p$ depending on whether the
real photon insertion was on the $p'$ or $p$ leg in the $n^{\rm th}$
order matrix element
${\cal M}_{n+1}$ (or its conjugate ${\cal M}_{n+1}^\dagger$). Again,
the insertion of a $\widetilde{K}$ real photon into an
$n^{\rm th}$ order graph leads to a cross section that is proportional
to the lower order one, with the cancellation again occurring pair-wise,
to give
\begin{align} \nonumber
\left \vert {\cal{M}}_{n+1}^{\widetilde{K}\gamma,{\rm tot}} \right \vert^2
	& \propto -e^2 \left[\strut \tilde{b}_k(p,p) -2
	\tilde{b}_k(p',p) + \tilde{b}_k(p',p') \right]~,
		\nonumber \\
  & \equiv -e^2 \widetilde{J}^2(k)~.
\label{eq:ktilde}
\end{align}
Thermal real photons can be both emitted into and absorbed from the heat
bath, in contrast to the case at zero temperature where only real photon
emission is permitted. This is accounted for in the phase space, with a
factor of 
\begin{equation}
\int \frac{{\rm d}^4 k}{(2\pi)^4} 2\pi \delta(k^2) \left[\theta(k^0) (1 +
n_B(\vert k^0 \vert)) + \theta(-k^0) n_B(\vert k^0 \vert) \right]~,
\label{eq:realphspace}
\end{equation}
for every real photon contribution. This reflects the fact that the
probability of emission of a real photon with momentum $k$ into a heat
bath at temperature $T$ is proportional to $(n_B(\vert k^0 \vert) +1)$
while the probability of absorbing a photon with the same momentum from
the heat bath is proportional to $n_B(\vert k^0 \vert)$. (The initial
flux remains the same as in the virtual photon case.)

The contributions from the virtual $K$ (Eq.~\ref{eq:K}) and real
$\widetilde{K}$ (Eq.~\ref{eq:ktilde}) photon corrections are IR
divergent and cancel order by order in the theory, so that the cross
section for the process $\chi \chi \to f \overline{f}$ is IR safe. We
are therefore left with the IR-finite $G$ photon contribution,
that we will now evaluate\footnote{Note that we have not computed the
$\widetilde{G}$ contribution which determines the cross section for
$\chi \chi \to f \overline{f} (\gamma)$, {\em i.e.}, real photon
emission/absorption.} in Sections~\ref{sec:cr_nlo_nondyn} and
\ref{sec:cr_nlo_dyn}. However, first, for completeness, we present the
LO results.

\section{The dark matter annihilation cross section at LO}
\label{sec:crlo}

The leading order (LO) contribution to the annihilation process $\chi
\overline{\chi} \to f \overline{f}$ arises from the $t$- and $u$-channel
processes shown in Fig.~\ref{fig:lo}. The momenta in the CM frame are
$q', q \to p', p$, with the choices
\begin{align}
q'^{\mu} & = (H, 0, 0, P)~, \qquad &
q^{\mu} & = (H, 0, 0, -P)~, \nonumber \\
p'^{\,\mu} & = (H, P'\sin\theta, 0, P'\cos\theta)~, \qquad &
p^{\mu} & = (H, -P'\sin\theta, 0, -P'\cos\theta)~,
\label{eq:kin}
\end{align}
where the centre of momentum energy, $s = 4H^2$, and $\theta$ is the angle
between the initial and final momenta ($\vec{q}', \vec{p}'$). The cross
section at leading order for this $2\to 2$ process is given as usual by,
\begin{eqnarray}
\sigma_{LO} & = & \frac{1}{64\pi^2 s}
~ \frac{\vert \vec{p'}\vert} {\vert \vec{q'}\vert}
\int \hbox{d}\Omega
\left[\sum_{spins}
~\vert {\cal{M}}_{LO}^t - {\cal{M}}_{LO}^u \vert^2 \right]~, \nonumber \\
 & = & \frac{1}{32\pi s}
~ \frac{P'} {P}
\int \hbox{d}\!\cos\theta
\left[\sum_{spins}
~\vert {\cal{M}}_{LO}^t - {\cal{M}}_{LO}^u \vert^2 \right]~,
\end{eqnarray}
where the integration over
the azimuth $\phi$ is trivial, and a summation over both
final state and initial state spins is to be performed since all
helicity configurations contribute to the total cross section.

The contribution from the matrix element can be expressed as
\begin{eqnarray}
\int \hbox{d}\!\cos\theta \sum_{spins} ~\vert {\cal{M}}_{LO}^t -
{\cal{M}}_{LO}^u \vert^2 
& \equiv & Int^t_{LO} + Int^u_{LO} - Int^{tu}_{LO}~, \nonumber \\
 & \equiv & Int_{LO}~,
 \label{eq:lo}
\end{eqnarray}
where the subscript $LO$ denotes the LO contribution and the three terms
on the right hand side refer to the square of the $t$-channel matrix
element, the square of the $u$-channel matrix element, and the $tu$ cross
terms respectively. Note that the cross term vanishes for Dirac-type
dark matter particles. The $t$-, $u$-channel matrix elements at LO
are given by,
\begin{align}
{\cal{M}}_{LO}^t & = 
{i \lambda ^2}
\left(\overline{v}(q,{m_\chi}) P_L \, v(p,{m_f})\right) \, \Delta(l)
\left(\overline{u}({p'},{m_f}) P_R \, u({q'},{m_\chi}) \right)~, \nonumber \\
{\cal{M}}_{LO}^u & =
{i \lambda ^2}
\left(\overline{v}(q',{m_\chi}) P_L \, v(p,{m_f})\right) \, \Delta(l')
\left(\overline{u}({p'},{m_f}) P_R \, u({q},{m_\chi}) \right)~, 
\label{eq:MLO}
\end{align} 
where we have lightened the notation for the scalar propagator so that
$\Delta(l) = i/(l^2-m_\phi^2)$, and similarly for $l'$, with
$l=q-p \equiv p' - q'$ and $l'=q' -p \equiv p'-q$ referring to
the momentum of the intermediate scalar for the $t$ and $u$ channels
respectively, and $P_{R,L} \equiv (1 \pm \gamma_5)/2$. Adding all the
contributions, we find,
\begin{align}
Int_{LO} & = \frac{2 \lambda ^4} {P\,P'(2 H^2-m_\Phi^2)}
\left\{-\left[H^2 \left(m_\chi^2-2
m_\Phi^2\right)+m_\chi^2 P'^2+m_\Phi^4\right]
\log \frac{\left(2 H^2-m_\Phi^2+2 P\,P'\right)}
{\left(2 H^2-m_\Phi^2-2 P\,P'\right)} \right. \nonumber \\
& \left. \qquad
+\frac{4 P\,P' \left(2 H^2-m_\Phi^2\right)
\left(2 H^4-2 H^2 m_\Phi^2+m_\Phi^4-2 P^2 P'^2\right)}{4
H^4-4 H^2 m_\Phi^2+m_\Phi^4-4 P^2\,P'^2} \right\}~,
\label{eq:LO}
\end{align}
where $m_\Phi^2 \equiv m_\chi^2 + m_f^2 - m_\phi^2$. The logarithmic
terms arise when either of $p, p'$ are
collinear with one of $q, q'$; however, these are not divergent;
they contribute at a single phase space point and are well-behaved. Then
the invariant cross section at LO can be written as
\begin{align}
\sigma_{LO}(s) & = \quad \frac{1}{32 \pi s} \frac{P'}{P} Int_{LO}~,
\label{eq:sigma_LO}
\end{align}
where $P'^2 = H^2 - m_f^2$, $P^2 = H^2 - m_\chi^2$, and $4 H^2 =
s$, the usual Mandelstam variable.

\paragraph{LO cross section in the non-relativistic limit}:
In the non-relativistic limit, when the velocity of the dark matter
particles is small, we can write $P = m_\chi v$ and $H^2 \approx
m_\chi^2(1+v^2)$. Since $P \ll H, m_\Phi$, we can expand the log term as
\begin{align}
\log \frac{\left(2 H^2-m_\Phi^2+2 P\,P'\right)}
{2 H^2-m_\Phi^2-2 P\,P'}
\overset{P~small}{\longrightarrow} 
2 \left[ \frac{2 P P'}{2 H^2-m_\Phi^2} + \frac{2}{3} 
\left(\frac{2 P P'}{2 H^2-m_\Phi^2}\right)^3 + \cdots
\right]~,
\label{eq:loge}
\end{align}
to get
\begin{align}
Int_{LO} & \overset{v~small}{\longrightarrow}
\quad \frac{8 \lambda^4 m_\chi^2 m_f^2}
{(m_\chi^2+ m_\phi^2 - m_f^2)^2} + {\cal{O}}(v^2)~,
\label{eq:LOv}
\end{align}
Notice that the cross section is proportional to the square of the fermion
mass, which is a well-known result \cite{Beneke:2014gla}.
Hence, the LO cross section can be written as
\begin{align}
\sigma_{LO}(s) & = \quad \frac{1}{32 \pi s} \frac{P'}{P} Int_{LO}~, \nonumber \\
 & \overset{v~small}{\longrightarrow} \frac{\lambda^4}{4 \pi s} \frac{P'}{P} 
\left[\frac{m_\chi^2 m_f^2} {(m_\chi^2+ m_\phi^2 - m_f^2)^2} + {\cal{O}}(v^2)
\right]~.
\label{eq:sigma_LOv}
\end{align}
This term is the usual velocity-independent ``$a$'' term in $s \,\sigma(s)
v_{rel} = a + b v^2$ for the annihilation process \cite{Beneke:2014gla};
with the relative velocity between the two dark matter particles given
by $v_{rel} = 2 v$, where $v$ is the CM velocity of either of the
particles. Terms of order ${\cal{O}}(v^2)$ can be calculated by
retaining higher orders in the expansion of Eq.~\ref{eq:LO}.
 
We can repeat the calculation in the limit when the scalar is
much heavier than the other particles, {\em viz.}, $m_\phi^2 \gg m_\chi^2
\gtrsim m_f^2$. Then $l^2 \equiv (q-p)^2 \ll m_\phi^2$ (where we have
implicitly assumed that $\sqrt{s} \ll m_\phi$) and the scalar
propagator can be approximated by $i D_\phi = i/(l^2-m_\phi^2)
\to -i/m_\phi^2$, so that we get
\begin{align}
Int_{LO}^{\rm heavy~scalar} & =
\frac{8 \lambda^4}{3 m_\phi^4}
\left[6 H^4-3 H^2 m_\chi^2+P'^2 (2 P^2-3 m_\chi^2)\right]~, \nonumber \\
 & = \frac{8 \lambda^4}{3 m_\phi^4}
 \left[8 H^2( H^2 - m_\chi^2) +m_f^2 (5 m_\chi^2 - 2 H^2)\right]~,
\end{align}
where we have substituted for $P, P'$ in the last line. This gives us a
cross section,
\begin{align}
\sigma_{LO}^{\rm heavy~scalar} & =
\frac{1}{12\pi s}
\frac{P'}{P}
\frac{\lambda^4}{m_\phi^4}
 \left[8 H^2( H^2 - m_\chi^2) +m_f^2 (5 m_\chi^2 - 2 H^2)\right]~.
\end{align}
The first term in the square brackets is proportional to $H^2 = m_\chi^2
\equiv P^2$ and the
second is proportional to $m_f^2$. In the non-relativistic limit, with
$H^2 = m_\chi^2(1+ v^2)$, $P= m_\chi v$, this matches the expression
given in Eq.~\ref{eq:sigma_LOv}, with the further approximation,
$(m_\chi^2+m_\phi^2-m_f^2) \to m_\phi^2$, which is valid in the heavy
scalar limit. The important features are instantly visible within this
approximation: the $1/m_\phi^4$ dependence arising from the scalar
propagator, and the $m_f^2$ dependence of the cross section in the
non-relativistic limit, $H\to m_\chi$. Therefore, we present the NLO
calculation of the cross section in the next sections in two parts: in
the next section, we present the results in the heavy-scalar
limit, where the expressions are shorter and the features can be easily
understood. In the subsequent section, we then present the results with
the fully dynamical scalar propagator. While the detailed results are
cumbersome and can be found on-line \cite{online}, the results in the
non-relativistic limit have been presented for comparison and discussion
here.

\section{The dark matter annihilation cross section at
NLO---``heavy-scalar'' approximation}
\label{sec:cr_nlo_nondyn}

We apply the Grammer and Yennie technique \cite{Grammer:1973db},
generalised for the case of thermal field theories as explained in
Section~\ref{sec:GY}. There are two sets of contributions at higher
order: insertions of a virtual photon in the LO diagrams shown in
Fig.~\ref{fig:lo}, as well as insertions of real photons which can
be both emitted into or absorbed from the heat bath at temperature
$T$.  Each of these contributions can be factorised into $K$ and $G$
(or $\widetilde{K}$ and $\widetilde{G}$) parts for virtual (or real)
photon insertions respectively, by applying the Grammer and Yennie
technique. As a consequence, the IR divergences are contained in the $K$
($\widetilde{K}$) contributions respectively and furthermore, were shown
earlier to cancel not only at NLO \cite{Beneke:2014gla}, but at all
orders \cite{Sen:2020oix} in perturbation theory. Note that only soft
photon emissions/absorptions of the process $\chi \chi \to f \overline{f}
(\gamma)$ are contained in the $\widetilde{K}$ contribution. The hard
real-photon emissions/absorptions that contribute to $\widetilde{G}$
photon insertions do not contribute to the process of interest
here. Hence, by applying the Grammer and Yennie technique, the NLO
corrections to $\chi \chi \to f \overline{f}$ can simply be calculated
by computing the $G$ terms of the virtual photon contributions alone.

\begin{figure}[btp]
\begin{center}
\includegraphics[width=0.32\textwidth]{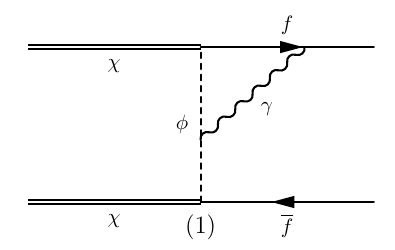}
\includegraphics[width=0.32\textwidth]{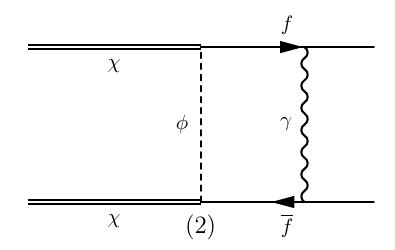}
\includegraphics[width=0.32\textwidth]{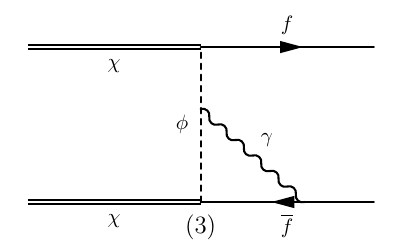}
\includegraphics[width=0.32\textwidth]{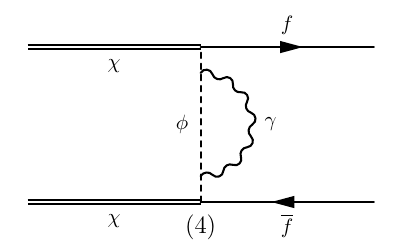}
\includegraphics[width=0.32\textwidth]{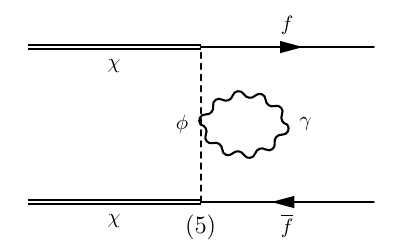}
\caption{\small \em The $t$-channel virtual photon corrections to the
dark matter annihilation process at next to leading order (NLO). Diagrams
are labelled from 1--5. Analogous contributions from the $u$-channel
diagrams exist.} 
\label{fig:nlo}
\end{center}
\end{figure}

The various contributions\footnote{Note that fermion self-energy
corrections do not contribute to the $G$ photon terms.} at NLO are
shown in Fig.~\ref{fig:nlo}. In order to compute the $G$ photon
contribution, we will replce the photon propagator by its $G$ photon
part; see Eq.~\ref{eq:KG}.  Furthermore, throughout this section, we
will approximate the scalar propagator by $i D_\phi = i/(l^2-m_\phi^2)
\to -i/m_\phi^2$, and lift this approximation in the next section.
Analogous to the LO cross section, we can write the next-to-leading order
(NLO) virtual contribution as
\begin{eqnarray}
\sigma_{NLO} & \propto & 
\left[\sum_{spins}
\left({\cal{M}}_{LO}^t - {\cal{M}}_{LO}^u
\right)^\dagger \left({\cal{M}}_{NLO}^t - {\cal{M}}_{NLO}^u \right) + h.c.
\right]~,
\label{eq:nlo}
\end{eqnarray}
where each of the higher order $t$-channel terms in Eq.~\ref{eq:nlo}
gets contributions from the five $t$-channel diagrams shown in
Fig.~\ref{fig:nlo} (and similarly, their $u$-channel counterparts).

We will discuss the details of the calculation using Diagram 1 as an
example, before we list the contributions from all the remaining diagrams.

\subsection{The NLO contribution from Diagram 1}
\label{sec:D1}

We start by discussing the pure $t$-channel NLO contribution from the
first diagram in Fig.~\ref{fig:nlo}. Since the scalar is very heavy, the
{\sf thermal} contribution from the scalar propagators, which contain
a factor $n_\phi = 1/(\exp[\beta E_\phi]-1)$, can be dropped since
$E_\phi \gtrsim m_\phi \gg m_\chi$ and $\beta m_\chi \equiv x \sim 20$
near freeze-out. Hence the scalar propagators can only be of (11) or
(22) type; see Appendix \ref{sec:feyn} for details on the {\sf thermal}
and {\sf temperature independent} parts of various propagators. Since all
external lines can only be of type-1 \cite{Altherr:1993tn}, and scalar
propagators of (12) or (21) type can be neglected, it can be seen that
{\em all} propagators are of type (11) and all vertices are of type-1
alone. In fact, we will find that this holds for all the contributions
at NLO, for all the diagrams.

Therefore, there are three possibilities for Diagram 1, {\em viz.},

\begin{enumerate}

\item the photon propagator (with momentum $k$) is {\sf thermal}, 

\item the fermion propagator (with momentum $p' +k$) is {\sf thermal},

\item or both are {\sf thermal},

\end{enumerate}
where, by {\sf thermal}, we refer to the explicitly $T$-dependent second
term in the propagators as defined in Appendix~\ref{sec:feyn}.  Since the
{\sf thermal} part of all propagators puts the particle on mass-shell,
the last option yields a term depending on the product $\delta(k^2)
\, \delta((p'+k)^2 - m_f^2)$. Since the external fermion is on-shell,
$p'^2 = m_f^2$; hence this product of delta functions requires $p'
\cdot k = 0$. But $\delta(k^2)$ implies $k^0 = \pm \vert \vec{k} \vert$
or the angle between $k$ and $p'$ should satisfy $\cos\theta_{kp'} = \pm
p'^0/\vert\vec{p'} \vert$ or $\vert \cos\theta_{kp'} \vert > 1$, which
is impossible, so there is no phase space available in the case that both
propagators are on-shell. Hence we only need to consider the contribution
of the first diagram when either one of the photon or the fermion propagator
is thermal. In particular, as discussed in Section~\ref{sec:GY}, we only
need to consider the IR-finite $G$-photon contribution with the
$g^{\mu\nu}$ term in the photon propagator being replaced by $G^{\mu\nu}$:
\begin{align} \nonumber -i
g^{\mu{\nu}} & \rightarrow -i \,
\left[\strut g^{\mu\nu} - b_k (p_f, p_i)\,k^\mu k^\nu \right] ~, \nonumber \\
& \equiv -i \, \left[\strut G_k^{\mu\nu}(p_f,p_i)\right]~.
\label{eq:KG1}
\end{align}
Since the virtual photon vertices in Diagram~1, Fig.~\ref{fig:nlo} are
on the final fermion leg and the scalar, we have $p_f=p', p_i=p$, with
$b_k(p_f, p_i)$ defined in Eq.~\ref{eq:bk}. We consider in turn, the
contribution from the {\sf thermal} parts of the photon and fermion
propagators respectively to the cross section for the process shown
in Diagram 1 of Fig.~\ref{fig:nlo}. We present the purely $t$-channel
contribution in detail in order to highlight some technicalities, and
then go on to the other contributions.

\subsubsection{NLO thermal photon contribution to Diagram 1}
\label{sec:1g}
The matrix element at NLO from the ($t$-channel) Diagram 1 in
Fig.~\ref{fig:nlo} when the $G$-photon propagator is {\sf thermal}, is
given by,
\begin{align}
{\cal{M}}^{t}_{NLO} (\hbox{\small Diagram}\, 1, \gamma) & = 
-\int \!\!\! \frac{\hbox{d}^4k} {(2\pi)^4}
\frac{i e^2 \lambda ^2}{2 k \cdot p'}
\left(2\pi \delta(k^2)n_B(\vert k^0\vert) \right) \Delta(l) \Delta(l+k)
\nonumber \\
 & \qquad \qquad \left[\left(\overline{u}({p'},{m_f})\gamma_\mu
\left(\slashed{k}+\slashed{p'}+m_f\right) P_R\, u({q'},{m_\chi}) \right)
\right. \nonumber \\
 & \qquad \qquad \qquad \left.
\left(k-2 p+2 q\right)_{\nu }
\left(\overline{v}(q,{m_\chi}) P_L \, v(p,{m_f}) \right)\
\right]
\ G_k^{\mu\nu}(p',p)~, \nonumber \\
& \equiv 
\int \frac{\hbox{d}^4k}{(2\pi)^4} \left(2\pi
\delta(k^2) n_B(\vert k^0 \vert) \right) F_{NLO}^{t,1\gamma}(k)~,
\label{eq:fig1pt}
\end{align}
where $n_B$ is the Bose distribution function given in Eq.~\ref{eq:nB}
and we have used the index $(\hbox{\small Diagram}\, 1,
\gamma)$ to indicate that this is the contribution from Diagram 1 of
Fig.~\ref{fig:nlo} when the photon is {\sf thermal}.
At this point, we are in a position to justify the simplification of the
scalar propagator in the heavy-scalar approximation. The scalar propagator
for the $t$-channel diagrams contains either the inverse of $(l^2 -
m_\phi^2)$, or $((l+k)^2 - m_\phi^2)$, with $l \equiv q -p \equiv
p'-q'$. Since $m_\chi \ll m_\phi$, $l^2 < m_\phi^2$. Since the photon
propagator is {\sf thermal}, $k^2=0$ and $l\cdot k \propto \vert k^0
\vert$ is small since a large $k^0 = \vert \vec{k}\vert$ will suppress
the contribution due to the presence of the $n_B(\vert k^0 \vert)$
term. Hence either of the scalar propagators can be approximated by just
the $(m_\phi^2)^{-1}$ terms.

A similar reasoning will hold in the next Section \ref{sec:1f} when we
consider the fermion propagator to be {\sf thermal}.  Then we can write
$(l+k)^2 = ((p'+k)-q')^2 \equiv (t-q')^2$ (see Eq.~\ref{eq:tsub}). Now
$(p'+k)^2 \equiv t^2=m_f^2$ since the fermion propagator is {\sf thermal}
and $t\cdot q$ should be small, otherwise the contribution will be killed
by the $n_F(\vert t^0\vert)$ term; see Eq.~\ref{eq:nF}. A similar argument
holds for Diagrams 2 and 3 in Fig.~\ref{fig:nlo} when the anti-fermion
is thermal, where we can write $(l+k)^2 = (q-p+k)^2 \equiv (q-t)^2$. For
the $u$-channel matrix elements, the scalar momentum is $l' = q'-p$ and
the same reasoning applies. Hence we will replace $\Delta(l) = i /(l^2 -
m_\phi^2) \to i/(-m_\phi^2)$, etc., in this section. When the scalar
is heavy, but not much larger than the dark matter particle or the
scale of the interaction ($\sqrt{s}$), we will reinstate these terms
and recompute the cross section in Section \ref{sec:cr_nlo_dyn}.

The integration over the photon momentum $k$ can be partly
completed using the delta function:
\begin{align}
\int \hbox{d}^4k \left(2\pi \delta(k^2)\right) F(k) & = 
2\pi \int_{-\infty}^{\infty} \hbox{d}k^0 \int_0^{\infty} K^2 \hbox{d}K
\int \hbox{d}\Omega_k \left[\delta((k^0)^2-K^2)\right] F(k^0, K,
\Omega_k)~, \nonumber \\
 & = 2 \pi \int \hbox{d}k^0 \int \hbox{d}\Omega_k
\int K^2 \hbox{d}K \frac{\left[ \delta(k^0 -K) + \delta(k^0+K)
\right]}{\vert 2k^0 \vert} F(k^0, K, \Omega_k)~, \nonumber \\
 & = \pi \int \hbox{d}\Omega_k
\left[\int_0^\infty \vert k^0 \vert \hbox{d}k^0 F(k^0, k^0, \Omega_k) +
\int_{-\infty}^0 \vert k^0 \vert \hbox{d}k^0 F(k^0, -k^0, \Omega_k) \right]~,
\nonumber \\
& \equiv \pi \int_0^\infty \omega \hbox{d}\omega \left[
\int \hbox{d}\Omega_k F_+(\omega, \omega, \Omega_k) + 
\int \hbox{d}\Omega_k F_-(-\omega, \omega, \Omega_k) \right]~,
\label{eq:kint}
\end{align}
where $K \equiv \vert \vec{k} \vert$. Hence there are two contributions
to ${\cal{M}}_{NLO}^{t}(\hbox{\small Diagram}\,1,\gamma)$ (and to each
such matrix element), one where $K \to k^0 \equiv \omega$, and the other
where $K \to -k^0 \equiv -\omega$, as can be seen from Eq.~\ref{eq:kint}.
Note that the lower limit ($\omega \to 0$) can be safely taken precisely
because the $G$ photon contribution is guaranteed to be IR-finite.
The purely $t$-channel {\sf thermal} photon contribution from Diagram 1 is
therefore given according to Eq.~\ref{eq:nlo} as
\begin{align}
\sigma_{NLO}^{t}(\hbox{\small Diagram}\,1,\gamma) & = \frac{1}{32\pi s(2\pi)^4}
~ \frac{\vert \vec{p'}\vert} {\vert \vec{q'}\vert}
\int \hbox{d}\cos\theta
\left[\sum_{spins}
\left({\cal{M}}_{LO}^{t} \right)^\dagger
{\cal{M}}_{NLO}^{t,1\gamma} + h.c.
\right]~, \nonumber \\
 & = \frac{1}{32 s(2\pi)^4}
~ \frac{\vert \vec{p'}\vert} {\vert \vec{q'}\vert}
\int \omega \hbox{d}\omega \, n_B(\omega) \left[
\int \hbox{d}\cos\theta \int \hbox{d}\Omega_k
\left[ F_+^{t,1,\gamma} + F_-^{t,1,\gamma} \right]
\right] ~, \nonumber \\
 & \equiv \frac{1}{32 s(2\pi)^4}
~ \frac{\vert \vec{p'}\vert} {\vert \vec{q'}\vert}
\int \omega \hbox{d}\omega \, n_B(\omega)
Int_{\hbox{\small Diagram}\, 1, \gamma}^{t}~.
\label{eq:sigma_1g}
\end{align}
On performing the angular integrations, the contribution of this term to the
NLO cross section is given by,
\begin{align}
Int^{t}_{\hbox{\small Diagram} \, 1,\gamma} & = 
 \frac{64 \pi e^2 \lambda ^4}{3 {m_\phi}^6 {P'}}
\left(4 {P'} \left(3 H^4+P^2 \ {P'}^2\right) +
\log \frac{H-{P'}}{H+P'} 3H \left(H^2+P^2\right) \left(H^2+{P'}^2\right)
\right)~.
\label{eq:Intt1g}
\end{align}
Note that the result in Eq.~\ref{eq:Intt1g} is independent of $\omega$;
in fact, the individual $F_+$ and $F_-$ contributions to
$Int_{\hbox{\small Diagram}\, 1, \gamma}^{t}$ contain some
apparent divergent terms of the order of
$1/\omega$. When the two terms are added, these cancel, leaving behind IR
finite terms that are integrable in $\omega$. This is a reflection of
the fact that the $G$ photon insertion was tailored precisely to yield
such an IR-finite result. The logarithmic terms arise
when $p'$ or $p$ are collinear with $q'$ or $q$. These will drop out of
the final calculation\footnote{The collinear terms cancel between the $G$
and $\widetilde{G}$ contributions from the virtual and real corrections
respectively \cite{Beneke:2014gla}.}.

Similarly, the $u$-channel and crossed $tu$-channel NLO contributions
from Diagram 1 of Fig.~\ref{fig:nlo} when the photon propagator is {\sf
thermal} can be calculated from the corresponding $u$-channel NLO matrix
element which is given by
\begin{align}
{\cal{M}}^{u}_{NLO} (\hbox{\small Diagram}\, 1, \gamma) & = 
- \int \!\!\! \frac{\hbox{d}^4k} {(2\pi)^4}
\frac{i e^2 \lambda^2}{2 k \cdot p'}
\left(2\pi \delta(k^2)n_B(\vert k^0\vert) \right)
\Delta(l') \Delta (l'+k) \nonumber \\
 & \qquad \qquad \left[\left(\overline{u}({p'},{m_f})\gamma_\mu
\left(\slashed{k}+\slashed{p'}+m_f\right)
P_R \, u(q,m_\chi) \right) \left(k-2 p+2 q'\right)_\nu \right. \nonumber \\
 & \qquad \qquad \left. \left(\overline{v}(q',m_\chi)
 P_L \, v(p,m_f) \right) \right] G_k^{\mu\nu}(p',p)~.
\label{eq:fig1pu}
\end{align}
Then the total NLO contribution from (the $t$-, $u$ and cross $tu$
channels of) Diagram 1, Fig.~\ref{fig:nlo}, with {\sf thermal} photon
propagator is given by,
\begin{align}
Int_{\hbox{\small Diagram}\, 1, \gamma}^{tt+uu-tu} & =
\frac{64 \pi e^2 \lambda ^4} {3 m_\phi^6 P'}
\left[24 H^4 P'-6 H^2 m_\chi^2
P' -6 m_\chi^2 P' \left(m_f^2+3 P'^2\right)+8 P^2 P'^3
\right. \nonumber \\
& \qquad \left. +\log \frac{H-P'}{H+P'} \left(6 H^5+H^3 \left(6
\left(P^2+P'^2\right)-3 m_\chi^2\right)+
3 H P'^2 \left(2 P^2-3 m_\chi^2\right)\right) \right]~, \nonumber \\
& = 
\frac{64 \pi e^2 \lambda ^4} {3 m_\phi^6 P'}
\left[4 P' \left(8 H^4-2 \ H^2 \left(4 m_\chi^2+m_f^2\right)+
5 m_\chi^2 m_f^2\right) \right. \nonumber \\
 & \left. \qquad +3 \log \frac{H-P'}{H+P'}
\left(8 H^5-4 H^3 \left(2 m_\chi^2+m_f^2\right)+5 H m_\chi^2 m_f^2\right)
\right]~, \nonumber \\
 & \overset{\rm non\hbox{-}coll}{\longrightarrow}
\frac{256 \pi e^2 \lambda ^4} {3 m_\phi^6}
\left(8 H^4-2 \ H^2 \left(4 m_\chi^2+m_f^2\right)+
5 m_\chi^2 m_f^2\right)~,
\label{eq:Int1tg_ttuutu}
\end{align}
where we have substituted for $P'^2=H^2-m_f^2$, $P^2=H^2-m_\chi^2$
and dropped the collinear logarithmic terms in the last line. It can be
seen that $Int_{\hbox{\small Diagram}\,1,\gamma}^{tt+uu-tu}$ given in
Eq.~\ref{eq:Int1tg_ttuutu} is independent of $\omega$; hence the final
contribution from {\sf thermal} photons to Diagram 1, Fig.~\ref{fig:nlo}
is
\begin{align}
\sigma_{NLO}^{tt+uu-tu}(\hbox{\small Diagram}\, 1, \gamma) & = 
 \frac{1}{32 s(2\pi)^4}
~ \frac{P'}{P}
\int \omega \hbox{d}\omega \, n_B(\omega)
Int_{\hbox{\small Diagram}\, 1, \gamma}^{tt+uu-tu}~, \nonumber \\
& = \frac{1}{32 s(2\pi)^4}
~ \frac{P'} {P}
~ \frac{\pi^2 T^2} {6} \times
Int_{\hbox{\small Diagram}\, 1, \gamma}^{tt+uu-tu}~,
\label{eq:crossNLO_t1g}
\end{align}
which is not only IR finite, as guaranteed by the Grammer and Yennie technique,
but has a $T^2$ temperature dependence of the cross section
due to the $\omega$-independence of $Int_{\hbox{\small
Diagram}\,1,\gamma}^{tt+uu-tu}$, with
\begin{align}
\int_0^\infty \omega \hbox{d}\omega \, n_B(\omega)
= \frac{\pi^2 T^2} {6}~.
\label{eq:omegaB_int}
\end{align}
We now discuss the contribution at NLO from Diagram 1,
Fig.~\ref{fig:nlo}, when the fermion propagator is {\sf thermal}. We
will see that not only this, but each contribution from the various
Diagrams in Fig.~\ref{fig:nlo} are IR finite, as a result of applying
the Grammer and Yennie technique.

\subsubsection{NLO thermal fermion contribution to Diagram 1}
\label{sec:1f}

As mentioned earlier, the thermal contributions from Diagram 1 only
arise when either the photon or fermion propagator is {\sf thermal},
that is, the explicitly $T$-dependent second term of the propagators in
Eqs.~\ref{eq:thermalD} and \ref{eq:fprop} contribute; the contribution
when both are {\sf thermal} vanishes. We now consider the contribution
at NLO of the $t$-channel Diagram 1, Fig.~\ref{fig:nlo}, when the
photon propagator is non-thermal, but the fermion propagator is {\sf
thermal}. Recall that only the photon contributions give rise to IR
divergences (both at $T=0$ and finite temperature). This is because
of their differing thermal distributions; see Eqs.~\ref{eq:nB} and
\ref{eq:nF}. In addition, we have shown in Section \ref{sec:thermalfermion}
that the $K$-photon contribution when the photon is inserted into a {\sf
thermal} fermion vanishes; see Eq.~\ref{eq:thermalfermionK}. Hence in
the $G$ photon insertion, which arises from $G^{\mu\nu} = (g^{\mu\nu}
- b_k k^\mu k^\nu)$, the second term, which is the $K$ photon piece,
vanishes and we therefore need consider just the $g^{\mu\nu}$ contribution
(or in other words, since the $K$ contribution vanishes, the $G$ term
contains the entire $g^{\mu\nu}$ contribution). The corresponding
matrix element is then given by
\begin{align}
{\cal{M}}^{t}_{NLO} (\hbox{\small Diagram}\, 1, f) & =
- \int \frac{\hbox{d}^4k} {(2\pi)^4}
\frac{i e^2 \lambda ^2 \, g^{\mu\nu}}{k^2} \,
\left(-2\pi \delta((p'+k)^2-m_f^2)n_F(\vert (p'+k)^0\vert) \right)
\nonumber \\
& \qquad\qquad
\Delta(l) \Delta(l+k)
\left[\left(\overline{u}({p'},{m_f})\gamma_\mu
\left(\slashed{k}+\slashed{p'}+m_f\right) P_R\, u({q'},{m_\chi}) \right)
\right. \nonumber \\
 & \qquad \qquad \left.
\left(k+2 p'-2 q'\right)_{\nu }
\left(\overline{v}(q,{m_\chi}) P_L \, v(p,{m_f}) \right)\
\right]~,
\label{eq:fig1ft}
\end{align}
where $n_F$ is the fermion number operator defined
in Eq.~\ref{eq:nF}, the $1/k^2$ term arises from the {\sf
temperature-independent} part of the photon propagator, and the
delta-function from the {\sf thermal} fermion propagator. As before, the
index $(\hbox{\small Diagram}\,1, f)$ refers to the contribution arising
from Diagram 1, Fig.~\ref{fig:nlo}, when the {\sf thermal} part of the
fermion propagator contributes. The sign
difference between $n_F$ and $n_B$ is what dictates the IR finite nature
of the fermionic contributions. It is convenient to define $p'+k \equiv
t$, and change the variable of integration to $t$, so that
\begin{align}
{\cal{M}}^{t}_{NLO} (\hbox{\small Diagram}\, 1, f) & =
\int \frac{\hbox{d}^4t} {(2\pi)^4}
\frac{i e^2 \lambda ^2}{2}
\frac{g^{\mu\nu}}{m_f^2 - t \cdot p'}\,
\left(2\pi \delta(t^2-m_f^2)n_F(\vert t^0\vert) \right)
\nonumber \\
& \qquad \Delta(l) \Delta(l+k)
\left[\left(\overline{u}({p'},{m_f})\gamma_\mu
\left(\slashed{t}+m_f\right) P_R\, u({q'},{m_\chi}) \right)
\right. \nonumber \\
 & \qquad \left.
\left(t+p'-2 q'\right)_{\nu }
\left(\overline{v}(q,{m_\chi}) P_L \, v(p,{m_f}) \right)\
\right]~,
\label{eq:tsub}
\end{align}
so that the same simplification can be done using the delta-function as
discussed in Eq.~\ref{eq:kint}, to obtain 
\begin{align}
\int \!\!\hbox{d}^4t \left[2\pi \delta(t^2\!\!-\!\!m_f^2)\right] \! F(t) & =
2\pi \int \hbox{d}^4t ~ \delta(t_0^2-\vert \vec{t} \vert^2 -m_f^2) F(t) 
\equiv 2\pi \int \hbox{d}^4t ~ \delta(t_0^2-\omega_t^2) F(t) 
~, \nonumber \\
& = \pi \int_{m_f}^\infty \! \! K_t 
~ \hbox{d}\omega_t \! \! \left[
\int \! \! \hbox{d}\Omega_t F_+(\omega_t, \omega_t, \Omega_t)
+ \! \! \int \hbox{d}\Omega_t F_-(-\omega_t, \omega_t, \Omega_t) \right],
\label{eq:Ot}
\end{align}
where $\omega_t^2 = \vert \vec{t}\vert^2 + m_f^2 \equiv K_t^2 + m_f^2$
and we have expressions for $F_\pm(t)$ analogous to that in
Eq.~\ref{eq:kint}. Then the {\sf thermal} fermion contribution to the
purely $t$-channel Diagram 1 is given analogously to that for {\sf
thermal} photons in Eq.~\ref{eq:sigma_1g} by
\begin{align}
\sigma_{NLO}^{t}(\hbox{\small Diagram}\,1,f) & = \frac{1}{32\pi s(2\pi)^4}
~ \frac{\vert \vec{p'}\vert} {\vert \vec{q'}\vert}
\int \hbox{d}\!\cos\theta
\left[\sum_{spins}
\left({\cal{M}}_{LO}^{t} \right)^\dagger
{\cal{M}}_{NLO}^{t,1 f} + h.c.
\right]~, \nonumber \\
 & = \frac{1}{32 s(2\pi)^4}
~ \frac{\vert \vec{p'}\vert} {\vert \vec{q'}\vert}
\int K_t \hbox{d}\omega_t \, n_F(\omega_t) \left[
\int \hbox{d}\!\cos\theta \int \hbox{d}\Omega_t
\left[ F_+^{t,1,f} + F_-^{t,1,f} \right]
\right] ~, \nonumber \\
 & \equiv \frac{1}{32 s(2\pi)^4}
~ \frac{\vert \vec{p'}\vert} {\vert \vec{q'}\vert}
\int K_t \hbox{d}\omega_t \, n_F(\omega_t)
Int_{\hbox{\small Diagram}\, 1, f}^{t}~.
\label{eq:sigma_1f}
\end{align}
Again, on completing the angular integration, we find the $t$-channel
{\sf thermal} fermion contribution to Diagram 1, Fig.~\ref{fig:nlo}, to be
\begin{align}
Int_{\hbox{\small Diagram}\, 1, f}^{t} & = 
\frac{32 \pi e^2 \lambda ^4} {3 K_t m_\phi^6 P'}
\left[4 K_t P' \left(3 H^2 m_\chi^2 +
P^2 \left(4 H^2 - m_f^2\right) \right) \right. \nonumber \\
 & \qquad -\left(6 H^5 \omega_t-3 H^4 m_f^2+ H^3 \omega_t \left(4 P^2-3 m_f^2\right)
-4 H^2 m_f^2 P^2-H \omega_t P^2 \left(m_f^2-2 P'^2\right)
\right. \nonumber \\
& \qquad - \left. 
m_f^2 P^2 (P'^2 - m_f^2)\right)
\log \frac{H \omega_t+K_t P'-m_f^2}{H \omega_t-K_tP'-m_f^2} \nonumber \\
& \qquad +\left(6 H^5 \omega_t+3 H^4 m_f^2+H^3 \omega_t
\left(4 P^2-3 m_f^2\right)
+4 H^2 m_f^2 P^2-H \omega_t P^2 \left(m_f^2-2 P'^2\right)
\right. \nonumber \\
 & \qquad + \left. \left. m_f^2 P^2 \left(P'^2-m_f^2\right)\right)
\log \frac{H \omega_t-K_t P'+m_f^2}{H \omega_t+K_t P'+m_f^2} \right]~.
\label{eq:Intt1f}
\end{align}
While the non-logarithmic terms are (as in the case of thermal photons)
independent of $\omega_t$, the logarithmic terms have a complicated
dependence on $\omega_t$ through $K_t$ and it is not possible to
analytically integrate these terms. A simplification is achieved in
the limit that the fermion masses can be neglected in comparison
to $\omega_t$, so that $K_t \to \omega_t$. Then the contribution
simplifies\footnote{Note that no approximations have been made in the
non-logarithmic term.} to
\begin{align}
Int_{\hbox{\small Diagram}\, 1, f}^{t} & \overset{K_t \to \omega_t}
{\longrightarrow} \frac{64 \pi e^2 H \lambda ^4} {3 {m_\phi}^6 {P'}}
\left[\rule{0pt}{16pt} 2 H {P'}
\left(3{m_\chi}^2 + 4 P^2\right) \right. \nonumber \\
 & \qquad \qquad \left. - 2 \left(3 H^4 + 2 H^2 P^2 + P^2 P'^2\right)
\log \frac{H +P'}{H-P'} \right]~,
\label{eq:Intt1f_mf0}
\end{align}
with
\begin{align}
\frac{\log [H \omega_t + K_t P' \pm m_f^2]}
{\log [H \omega_t - K_t P' \pm m_f^2]} & \overset{K_t\to
\omega_t}{\longrightarrow} \frac{\log [H + P']}{\log [H - P']}~,
\label{eq:log}
\end{align}
where we have dropped terms of order ${\cal{O}}(m_f^2/(H\omega_t))$, so
that these logarithmic terms are of the same form as before, independent
of $\omega_t$, yielding again a $T^2$ temperature dependence from this
contribution, using\footnote{Expanding $K_t = \sqrt{\omega_t^2 - m_f^2}$
as $K_t \approx \omega_t - m_f^2/(2\omega_t) + \cdots$ is not a possible
choice since $\int( {\rm d} \omega_t /\omega_t) n_F$ is not convergent;
it is of course possible to integrate this numerically if more precise
results are required.}
\begin{align}
\int_{m_f}^\infty K_t \hbox{d}\omega_t \, n_F(\omega_t) \approx
\int_0^\infty \omega_t \hbox{d}\omega_t \, n_F(\omega_t)
= \frac{\pi^2 T^2} {12}~.
\label{eq:omegaF_int}
\end{align}
Similarly, the $u$-channel matrix element for thermal fermions is given
by:
\begin{align}
{\cal{M}}^{u}_{NLO} (\hbox{\small Diagram}\, 1, f) & =
- \int \frac{\hbox{d}^4t} {(2\pi)^4}
\frac{i e^2 \lambda ^2
\,g^{\mu\nu}}{2(m_f^2 - t \cdot p')}\,
\left(-2\pi \delta(t^2-m_f^2)n_F(\vert t^0\vert) \right)
\nonumber \\
& \qquad \qquad
\Delta(l') \Delta (l'+k)
\left[\left(\overline{u}({p'},{m_f})\gamma_\mu
\left(\slashed{t}+m_f\right) P_R\, u({q},{m_\chi}) \right) \right.
\nonumber \\
& \qquad \qquad \left.
 \left(t+p'-2 q\right)_{\nu }
\left(\overline{v}(q',{m_\chi}) P_L \, v(p,{m_f}) \right)\right]~.
\label{eq:fig1fu}
\end{align}
Using the LO matrix elements given in Eq.~\ref{eq:MLO}, the NLO matrix
element for the $t$-channel contribution given in Eqs.~\ref{eq:fig1ft}
and \ref{eq:tsub} and that for the $u$-channel contribution in
Eq.~\ref{eq:fig1fu}, the total thermal fermion contribution from
Diagram 1, Fig.~\ref{fig:nlo}, that is, from the combined $t$-channel,
$u$-channel, and the crossed $tu$-channel, is found to be
\begin{align}
Int_{\hbox{\small Diagram}\,1,f}^{tt+uu-tu} & =
\frac{32 \pi e^2 \lambda ^4} {3 m_\phi^6 K_t P'}
\left[4 K_t P'\left(3 m_\chi^2 m_f^2 +
2P^2(4 H^2 -m_f^2) \right) \right. \nonumber \\
& \qquad +3 m_\chi^2 \log \frac{H \omega_t + K_t P' - m_f^2}
{H \omega_t - K_t P' - m_f^2} \left( 3H^3\omega_t - 2 H^2 m_f^2
+ H \omega_t (P'^2 - 2 m_f^2) + m_f^4 - P'^2 m_f^2\right) \nonumber \\
& \qquad -3 m_\chi^2 \log \frac{H \omega_t + K_t P' + m_f^2}
{H \omega_t - K_t P' + m_f^2} \left(3H^3\omega_t + 2 H^2 m_f^2 + H
\omega_t (P'^2 - 2 m_f^2) - m_f^4 + P'^2 m_f^2\right) \nonumber \\
& \qquad -2 \log \frac{H \omega_t + K_t P' - m_f^2}
{H \omega_t - K_t P' - m_f^2} \left( 6H^5\omega_t - 3 H^4 m_f^2 + H^3
\omega_t (4P^2 - 3 m_f^2) - 4H^2P^2 m_f^2 \right. \nonumber \\
& \qquad \qquad \qquad \qquad \left. - H \omega_t P^2(m_f^2 - 2
P'^2) - P^2 m_f^2( P'^2- m_f^2)\right) \nonumber \\
& \qquad -2 \log \frac{H \omega_t + K_t P' + m_f^2}
{H \omega_t - K_t P' + m_f^2} \left(6H^5\omega_t + 3 H^4 m_f^2 + H^3
\omega_t (4P^2 - 3 m_f^2) + 4H^2P^2 m_f^2 \right. \nonumber \\
& \qquad \qquad \qquad \qquad \left. \left. - H \omega_t P^2(m_f^2 - 2
P'^2) + P^2 m_f^2( P'^2- m_f^2)\right) \right]~.
\label{eq:Int1f_all}
\end{align}
As in the case of the $t$-channel contribution alone, we can again use
the approximation given in Eq.~\ref{eq:log}, so that
\begin{align}
Int_{\hbox{\small Diagram}\, 1, f}^{tt+uu-tu} &
\overset{K_t\to \omega_t}{\longrightarrow}
\frac{128 \pi e^2 \lambda ^4} {3 m_\phi^6 }
\left[\left( 3 m_\chi^2 m_f^2 + 2P^2(4 H^2 -m_f^2) \right)
-\frac{3 H}{2 P'} \log \frac{H+P'}{H-P'}
\left\{4 H^4+ \right. \right. \nonumber \\
 & \qquad \qquad \qquad \qquad \left. \left.
 2 H^2 \left(8 P^2-2 m_\chi^2 - m_f^2\right) + m_f^2
\left(3 m_\chi^2 - 2 P^2 \right) \right\} \rule{0pt}{16pt}
\right]~.
\label{eq:1f}
\end{align}
Again $Int_{NLO}^{1,tt+uu-tu,f}$ is independent of $\omega_t$; hence,
the $\omega_t$ integral (see Eq.~\ref{eq:omegaF_int}) gives a
$T^2$ temperature dependence to the cross section (analogue of
Eq.~\ref{eq:sigma_1f} for all channels) from the {\sf thermal} fermions
in Diagram 1, Fig.~\ref{fig:nlo}, as well.

\subsection{Total thermal contribution to the NLO cross section}

Now that we have demonstrated details of the calculation of the cross
section from Diagram 1, Fig.~\ref{fig:nlo}, we present the detailed
results from all Diagrams in Fig.~\ref{fig:nlo}. Diagram 3 is similar
to Diagram 1; here also, it is clear from an analysis similar to that
for Diagram 1 in Section~\ref{sec:D1} that only one of the photon and
fermion propagators can be {\sf thermal} at a time. Note that Diagrams 4
and 5 have no fermion propagators while Diagram 2 has three propagators
(excluding that of the scalar) and can in principle have contributions
from any one, any two, or all three propagators being {\sf thermal}
(where we remind the reader that by {\sf thermal} we mean the contribution
from the explicitly $T$-dependent second term in the expressions for
the photon and fermion propagators given in Eqs.~\ref{eq:thermalD}
and \ref{eq:fprop}). As argued for Diagrams 1 and 3, there is no
consistent available kinematic phase space when the photon propagator
and either one of the fermion or anti-fermion propagators are both
{\sf thermal}. Hence there is no contribution from the case when all
three propagators in Diagram 2 are {\sf thermal} (all three propagators
contribute on-shell) as well as when the photon propagator and either
the fermion or anti-fermion propagator is {\sf thermal}. Left is the
case when the fermion and anti-fermion propagators are {\sf thermal},
but not the photon one. Here, we find that the delta function constraint
$\delta((p'+k)^2 - m_f^2) \delta((p-k)^2 - m_f^2)$ is satisfied for the
single phase space point, $k^0=0$. Therefore, for all the diagrams, as
long as we discard the negligible contribution from scalar {\sf thermal}
propagators, we only have thermal contributions when exactly one of the
propagators in the diagram is {\sf thermal} and the others contribute
through their non-thermal or {\sf temperature-independent} parts.

We now go on to present the contributions from the remaining Diagrams
2--5 in Fig.~\ref{fig:nlo} as well as their $u$-channel
counterparts. As can be observed from the diagrams in
Fig.~\ref{fig:nlo} itself, Diagrams 1, 3 and 5 have two scalar propagators,
Diagram 4 has three, and Diagram 2 as well as the LO matrix element have
one scalar propagator each. Hence it is expected that the contributions
of Diagram 4 will be suppressed compared to the other contributions.
As before, we have only one propagator whose thermal part contributes,
and we separately list the contributions from the terms when the photon
propagator is {\sf thermal} and when the fermion/anti-fermion one is
{\sf thermal}. We begin with the {\sf thermal} photon contributions.

\subsubsection{Total thermal photon contributions to the cross section}

The NLO matrix elements for the remaining Diagrams 2--5,
Fig.~\ref{fig:nlo}, when the thermal part of the photon propagator
is taken into account are analogous to the expression given in
Eq.~\ref{eq:sigma_1g} and are given in Appendix ~\ref{sec:appb}. The
corresponding contributions from Diagrams 2--5 to the cross section
(corresponding to Eqs.~\ref{eq:sigma_1g}, \ref{eq:Int1tg_ttuutu},
\ref{eq:crossNLO_t1g} from Diagram 1) from both the $t$- and $u$-channels
and crossed $tu$-channel (in the heavy scalar propagator limit)
are given by,
\begin{align}
Int_{\hbox{\small Diagram}\,2,\gamma}^{tt+uu-tu} & =
 \frac{64 \pi e^2 \lambda ^4}{3 m_\phi^4} \left[
 \left(6 H^2-6 m_\chi^2+2 P^2\right) + 
\frac{1}{HP'} \log \frac{H\!\!-\!\!P'}{H\!\!+\!\!P'}
 \left(3 H^2\!\!-\!\!P^2\right) \ (H^2\!\!-\!\!P'^2) \right]~,
\end{align}
\begin{align}
Int_{\hbox{\small Diagram}\,3,\gamma}^{tt+uu-tu} & =
 \frac{64 \pi e^2 \lambda ^4} {3 m_\phi^6 P'}
 \left[2 P' \left(6 H^4+H^2 \left(8 P^2-3 m_\chi^2\right)
 +\left(3 m_\chi^2-2 P^2\right)
\left(m_f^2-P'^2\right)\right) \right. \nonumber \\
& \qquad +H \log \frac{H-P'}{H+P'}
\left(12 H^4+H^2 \left(-9 m_\chi^2-6 m_f^2+8 P^2\right) \right. \nonumber \\
& \qquad \left. \left. +m_\chi^2
\left(6 m_f^2-3 P'^2\right)-2 P^2 \left(m_f^2-2 P'^2\right)\right) \right]~,
\end{align}
\begin{align}
Int_{\hbox{\small Diagram}\,4,\gamma}^{tt+uu-tu} & =
\frac{512 \pi e^2 \lambda ^4}
{15 m_\phi^8 P' (H^2-P'^2)}
\left[P' \left\{60 H^8-30 H^6 \left(2 m_\chi^2-m_f^2+2 P'^2\right)
\right. \right. \nonumber \\
& \qquad +15 H^4 \left(m_\chi^4-m_\chi^2 \left(m_f^2-2 P'^2\right)
+2 P^2 \left(3 m_f^2+2 P'^2\right)\right)\nonumber \\
& \qquad -H^2 \left[5 m_\chi^2 \left(m_f^2 \left(P^2+3 P'^2\right)
+2 P'^2 \left(P^2-3 P'^2\right) \right) \right. \nonumber \\
& \qquad \left. +2 P^2 \left(m_f^2
\left(15 P'^2-4 P^2\right)+30 P'^4\right)\right]\nonumber \\
& \qquad \left. -P'^2 \left(15 m_\chi^4 P'^2+5 m_\chi^2 P^2
\left(m_f^2-2 P'^2\right)+2 m_f^2 P^4\right)\right\} \nonumber \\
& \qquad + \log \frac{H-P'}{H+P'} H m_f^2 \left(H^2-P'^2\right)
\left\{30 H^4-5 H^2 \left(3 m_\chi^2-8 P^2\right) \right. \nonumber \\
 & \qquad \left. \left. -15 m_\chi^2 P'^2+4 P^4+10 P^2 P'^2\right\}
\right]~,
\end{align}
\begin{align}
Int_{\hbox{\small Diagram}\,5,\gamma}^{tt+uu-tu} & =
\frac{512 \pi e^2 \lambda^4} {3 m_\phi^6}
\left(-6 H^4+3 H^2 m_\chi^2+P'^2 \left(3 m_\chi^2-2 P^2\right)\right)~.
\end{align}
Simplifying, and substituting $P'^2 \to H^2 - m_f^2$ and $P^2 \to H^2 -
m_\chi^2$, and dropping the collinear terms, the total contribution
from all terms where the virtual photon propagator is thermal, is given by
\begin{align}
Int_{Total,\gamma T}^{tt+uu-tu} & =
\left\{ \left[Int_{\hbox{\small Diagram}\,2,\gamma}^{tt+uu-tu}\right] +
\left[Int_{\hbox{\small Diagram}\,1,\gamma}^{tt+uu-tu} +
Int_{\hbox{\small Diagram}\,3,\gamma}^{tt+uu-tu} +
Int_{\hbox{\small Diagram}\,5,\gamma}^{tt+uu-tu}\right] \right. + \nonumber \\
& \left. \qquad \qquad \left[Int_{\hbox{\small Diagram}\,4,\gamma}^{tt+uu-tu} \right]
\right\}~, \nonumber \\
 & = \frac{128 \pi e^2 \lambda ^4} {3} \left\{ \frac{1}{m_\phi^4}
\left[4 P^2\right] - \frac{2}{m_\phi^6} H^2 \left[H^2-m_\chi^2-P^2 \right]
+ \frac{4}{5 m_\phi^8} \left[90 H^6-120 H^4(m_\chi^2-P^2)\right. \right.
\nonumber \\
 & \qquad + H^2(30 m_\chi^4 + 5 m_\chi^2(9 m_f^2 - 4 P^2)
- 6 P^2(5 m_f^2 - P^2)) \nonumber \\
 & \qquad \qquad\qquad \left. \left. +m_f^2(-15 m_\chi^4 +
 15 m_\chi^2 P^2 + 2 P^4) \right] \rule{0mm}{8mm} \right\}~, \nonumber \\
 & = \frac{512 \pi e^2 \lambda ^4} {15} \left\{ \frac{5}
 {m_\phi^4} \left( H^2 - m_\chi^2\right) + \frac{H^2}{m_\phi^8} \left[
216 H^4- 272 H^2 m_\chi^2 + 56 m_\chi^4 \right] \right. \nonumber \\
 & \left. \qquad \qquad \qquad -\frac{m_f^2 }{m_\phi^8}\left[ 28 H^4 - 86 H^2 m_\chi^2 +
 28 m_\chi^4 \right] \right\}~,
\label{eq:Intgnd}
\end{align}
where we have combined contributions having the same number of
scalar propagators. A few points can be noted. From the first line,
we see that the total thermal photon contribution from Diagrams 1, 3,
and 5 vanishes, since $P^2 = H^2 - m_\chi^2$. Hence terms
contributing as $1/m_\phi^6$ from thermal photon insertions vanish. Also,
from the last line in Eq.~\ref{eq:Intgnd}, we see that the contribution
from Diagram 2 (contributing as $1/m_\phi^4$) is proportional to $H^2 -
m_\chi^2 = P^2$ and is small in the non-relativistic limit. So also
is the contribution from Diagram 4 ($1/m_\phi^8$) term since the
term proportional to $H^2$ also vanishes when $H = m_\chi$ (exact
non-relativistic limit), leaving only the $m_f^2$ dependent term.
The familiar independence on $\omega$ is seen for the total thermal
photon contribution at NLO, so that the total cross section from these
terms again has a $T^2$ temperature dependence. We now calculate the
contribution from thermal fermions.

\subsubsection{Total thermal fermion contributions to the cross section}

Only Diagrams 1--3 in Fig.~\ref{fig:nlo} contribute to the NLO cross
section when the {\sf thermal} part of the fermion propagator is considered. The
corresponding matrix elements for Diagrams 2 and 3 are also given in
Appendix \ref{sec:appb}. The contributions to the NLO cross section from
each of these diagrams (in the heavy scalar limit, where we
have substituted for $P'$ and dropped the logarithmic terms, which have
the same form as for Diagram 1), are given by
\begin{align}
Int_{\hbox{\small Diagram}\,2, f}^{tt+uu-tu} & =
-\frac{32 \pi e^2 \lambda^4}{3 m_\phi^4 \left(H^2 - \omega_t^2\right)}
\left(12 H^2(H^2 - m_\chi^2) +
4 H^2 P^2 + 2 \omega_t^2 (3 (H^2-m_\chi^2) + P^2) \right)~, \nonumber \\
Int_{\hbox{\small Diagram}\,2, \overline{f}}^{tt+uu-tu} & =
Int_{\hbox{\small Diagram}\,2, f}^{tt+uu-tu}~, \nonumber \\
Int_{\hbox{\small Diagram}\,3,\overline{f}}^{tt+uu-tu} & =
\frac{128 \pi e^2 \lambda^4}{3 m_\phi^6}
\left( 3 m_\chi^2m_f^2 + 2 P^2(4H^2- m_f^2) \right)~.
\end{align}
Notice that the contribution of the {\sf thermal} fermion from Diagram
3 is the same as that from Diagram 1, as can be seen\footnote{Note the
absence of odd powers of $\omega_t$ due to the symmetry explicit in
Eq.~\ref{eq:K}.} from Eq.~\ref{eq:1f}. Furthermore, the contribution of
the {\sf thermal} fermion from Diagram 2, Fig.~\ref{fig:nlo} has non-trivial
$\omega_t$ dependences both in the numerator and denominator. Recognising
that large values of $\omega_t$ will lead to vanishing of the
corresponding distribution function, $n_F(\vert \omega_t\vert)$, we
expand the denominator for $\omega_t < H$ (recall that $H \ge m_\chi$ and
$\beta m_\chi = x \sim 20$ at freeze-out) to obtain the total {\sf
thermal} fermion contribution from Diagram 2 to be
\begin{align}
Int_{\hbox{\small Diagram}\,2, f+\overline{f}}^{tt+uu-tu} & =
-\frac{64 \pi e^2 \lambda^4}{3 m_\phi^4 H^2}
\left(12 H^2(H^2 - m_\chi^2) +
4 H^2 P^2 + 6 \omega_t^2 (3 (H^2-m_\chi^2) + P^2) \right)~.
\label{eq:nlo_2ffbar}
\end{align}
The presence of the $\omega_t^2$ terms in the numerator will lead to
$T^4$ temperature dependence in the cross section\footnote{Since
$H^2 - m_\chi^2 = p^2$, both the $\omega_t^0$ and the $\omega_t^2$ terms
are small in the non-relativistic limit.}. Then, the total contribution
from all terms where the virtual fermion (or anti-fermion) propagator
is {\sf thermal}, is given by
\begin{align}
Int_{Total,(f+\overline{f}) T}^{tt+uu-tu} & =
\left[Int_{\hbox{\small Diagram}\,1, f}^{tt+uu-tu} +
Int_{\hbox{\small Diagram}\,3, \overline{f}}^{tt+uu-tu} \right] +
\left[Int_{\hbox{\small Diagram}\,2, f}^{tt+uu-tu} +
Int_{\hbox{\small Diagram}\,2, \overline{f}}^{tt+uu-tu} \right]
~, \nonumber \\
 & = \frac{64 \pi e^2 \lambda^4}{3 m_\phi^6} \left\{
\left[\rule{0pt}{14pt} 12 m_\chi^2m_f^2 + 8 P^2(4H^2- m_f^2) \right] -m_\phi^2
 \left[ \rule{0pt}{14pt} 12 (H^2 - m_\chi^2) + 4 P^2 \right.\right. \nonumber \\
& \qquad\qquad \left. \left. + \frac{6 \omega_t^2}{H^2} 
\left(3 (H^2-m_\chi^2) + P^2 \right) \right] \right\}~.
\label{eq:Intfnd}
\end{align}
The total invariant NLO cross section is then given by the sum of the
{\sf thermal} photon and {\sf thermal} fermion contributions, 
\begin{align}
\sigma_{NLO}(s) & = 
\frac{1}{32 s(2\pi)^4} ~ \frac{P'}{P} \left[\int \omega {\rm d} \omega
\, n_B Int_{Total,\gamma T}^{tt+uu-tu} + \int K_t {\rm d}\omega_t \, n_F
Int_{Total,(f+\overline{f}) T}^{tt+uu-tu} \right]~,
\nonumber \\
& = \frac{\lambda^4}{12 \pi s} ~ \frac{P'}{P} ~ \left[\frac{8 \alpha
T^2} {15} \right] \times 
 \left\{ \rule{0mm}{6mm}
\frac{1}{m_\phi^4}\left[ \frac{7 \pi^2 T^2}{4 H^2} (H^2 - m_\chi^2) \right] +
\frac{1}{m_\phi^6} \left[ \rule{0pt}{12pt} 5 m_\chi^2m_f^2 + 2 P^2(4H^2-
 m_f^2) \right] \right. \nonumber \\
 & \qquad \left.  + \frac{1}{m_\phi^8} \left[ \rule{0pt}{14pt} 8 H^2
(27 H^4- 34 H^2 m_\chi^2 + 7 m_\chi^4) - 2 m_f^2 (14 H^4 - 43 H^2 m_\chi^2 +
 14 m_\chi^4) \right] \right\}~;
\label{eq:crossNLO}
\end{align}
here $s = 4H^2$ and the terms have been ordered in increasing powers of
$m_\phi^2$ in the denominator. We have replaced $K_t$ by $\omega_t$ in
the integrand of the second term in order to obtain an analytical result
and furthermore, presented the results, taking the lower limit of the
integration to be zero. Notice that the leading term ($1/m_\phi^4$) is
of order ${\cal{O}}(T^4)$ and is further suppressed by $(H^2 - m_\chi^2)=
P^2$ in the non-relativistic limit. The terms suppressed by an additional
propagator factor $(\sim 1/m_\phi^6)$ contribute at $T^2$ order and are
either proportional to $m_f^2$ ($ \sim 5 m_\chi^2 m_f^2$) or to $P^2$,
which is small is the non-relativistic limit. However, these terms,
of order ${\cal{O}}(H^2 P^2 T^2/m_\phi^6)$ or ${\cal{O}}(m_\chi^2 m_f^2
T^2/m_\phi^6)$, can have significant contribution in the relativistic
regime when $P$ is large.

\paragraph{The NLO cross section in the non-relativistic limit}:
Freeze-out occurs around $m_\chi/T \sim 20$ so that the dark matter
particles can be considered to be non-relativistic at this point,
with $P\to m_\chi v$, and $H^2 \to m_\chi^2(1 + v^2)$, so that
Eqs.~\ref{eq:Intgnd} and \ref{eq:Intfnd} reduce to
\begin{align}
Int_{Total,\gamma T}^{tt+uu-tu}
 & \overset{v~small}\longrightarrow \frac{512 \pi e^2 \lambda ^4}
 {15 {m_\phi}^8} \left\{
 30 m_\chi^4 m_f^2 (1+ v^2) + 5 m_\phi^4 m_\chi^2 v^2 \right\}~,
 \nonumber \\
Int_{Total,(f+\overline{f}) T}^{tt+uu-tu} & \overset{v~small}\longrightarrow
\frac{64 \pi e^2 \lambda^4}{3 m_\phi^6} \left\{
\left[ 6 m_\chi^2 m_f^2 + 4 m_\chi^2 v^2 (4 m_\chi^2 - m_f^2) \right]
+ 8 m_\phi^2 m_\chi^2 v^2 \left(3 \frac{\omega_t^2}{H^2} -
2 \right) \right\}~.
\label{eq:NLOIntv}
\end{align}
It can be seen that the NLO cross section (obtained from
Eqs.~\ref{eq:crossNLO} and \ref{eq:NLOIntv}) is proportional to $m_f^2$
as $v\to 0$, just as the LO cross section; see Eq.~\ref{eq:LOv}. This can
be understood from helicity conservation: just as in the LO case, the NLO
diagrams (with an additional virtual photon) that we have computed
are planar $2\to 2$ processes. The Majorana coupling then forces the
final states into the ``wrong'' chirality so that the cross section is
proportional to the fermion mass squared. At early times in the evolution
of the Universe, the dark matter particles are relativistic, $\sqrt{s}$
can be large, and there is less suppression in the annihilation into the
lighter fermions. This also holds for freeze-in scenarios where $m_\chi/T
\sim {\cal{O}}(1)$. Now that we have understood the structure of the NLO
contributions in the heavy-scalar limit, we will go on to include the
entire scalar propagator in the calculation of the NLO cross section.

\section{The dark matter annihilation cross section at
NLO including the ``dynamical'' scalar propagator}
\label{sec:cr_nlo_dyn}

We repeat the calculation of the previous section without making the
heavy scalar assumption, {\em i.e.}, retaining the ``dynamical''
scalar propagator, $iD_\phi = i/(l^2 - m_\phi^2)$, where $l =
q -p$ is the momentum of the scalar. As we had already argued in Section
\ref{sec:1g}, $l \cdot k, k^2 \ll m_\phi^2$; hence, we expand the relevant
propagator terms, for instance,
\begin{align}
\frac{1}{(l+k)^2 - m_\phi^2} & =
\frac{1}{(l^2 - m_\phi^2) + (2 l\cdot k + k^2)} \nonumber \\
 & \approx \frac{1}{(l^2 - m_\phi^2)} \left[1 - \frac{(2 l\cdot k + k^2)}
 {(l^2 - m_\phi^2)} \right]~.
\end{align}
Similar expansions can be obtained for the $u$-channel scalar propagators
as well. This enables us to get analytic expressions for the NLO cross
section. However, the results are rather cumbersome and opaque; they are
available as Mathematica \cite{Mathematica} Notebooks in Ref.~\cite{online}. We simply make
a few remarks on the calculation and specifically on the non-relativistic
limit which is of interest at freeze-out.

The main difference on including the momentum dependence of the scalar
is the increase in complexity of the $\cos\theta$ angular integration
due to the presence of the term $l^2 - m_\phi^2 \equiv m_\chi^2 + m_f^2 -
m_\phi^2 - 2 q \cdot p \equiv m_\Phi^2 - 2(H^2 - P P' \cos\theta)$ in
the denominator. It will be seen that this gives rise to logarithms of
the type $\log\left(2 H^2-m_\Phi^2 \pm 2 P\,P'\right)$ which we have
already encountered when calculating the LO cross section.

Once again, the leading ${\cal{O}}(T^2)$ terms are $\omega$-
($\omega_t$)-independent for the {\sf thermal} photon (fermion) cases, while
the $\omega^2$ ($\omega_t^2$) terms contribute to the ${\cal{O}}(T^4)$
terms. We have,
\begin{align}
\sigma_{NLO}(s) & = \quad \frac{1}{32 s (2\pi)^4} \frac{P'}{P}
\left[ \int {\rm d} \omega\omega \, n_B 
~ Int_{Total,\gamma T}^{tt+uu-tu} +
\int {\rm d} \omega_t \omega_t \, n_F 
~ Int_{Total,(f+\overline{f}) T}^{tt+uu-tu} \right]~,
\nonumber \\
 & \overset{v~small}{\longrightarrow} \frac{\lambda^4}{4 \pi s} \frac{P'}{P}
\frac{\pi \alpha T^2}{6} \left[Int_{NLO}^a + v^2 \, Int_{NLO}^b \right]~,
\label{eq:sigma_NLOv}
\end{align}
where the expression in the second line refers to the so-called $a$ and $b$
terms that contribute in the non-relativistic limit.

It is interesting to note that the pattern is the same as with the heavy
scalar computation in the previous section. The leading behaviour from
naive counting of the scalar propagator is expected to have the form
(from Diagram 2)
\begin{align}
\sigma_{NLO}(s) \propto \frac{m_\chi^2 T^2}
{(2 H^2 - m_\Phi^2 + 2 P P') (2 H^2 - m_\Phi^2 - 2 P P')}~,
\end{align}
where the denominator (proportional to $m_\phi^4$ in the heavy scalar
limit) is the analogous term to the $1/m_\phi^4$ contribution in the
heavy-scalar limit. However, as in the previous section with heavy-scalar
assumption, we find this term vanishes and the leading contribution
is again
\begin{align}
\sigma_{NLO}(s) & \propto \frac{m_\phi^2 M^2 m_f^2 T^2}
    {(4 H^4 - 4 m_\Phi^2 H^2 + m_\Phi^4 - 4 P^2 P'^2)^2}~, \nonumber \\
 & \propto \frac{M^2 m_f^2 T^2}{m_\phi^6}~,
\end{align}
where $M^2$ is any combination of $H \equiv \sqrt{s}/2, m_\chi$ or
$m_f$, of dimension 2.

As mentioned earlier, the results for the general relativistic case are
available as sets of Mathematica Notebooks at the web-page
Ref.~\cite{online}. Here we will discuss some aspects of the results in
the non-relativistic limit, which are of relevance near freeze-out, when
$m_\chi/T \sim 20$.

\paragraph{NLO cross section in the non-relativistic limit}:

In the limit when $p \to m_\chi v$, and the expressions are expanded
for small velocities, the cross section can be expressed as $\sigma
\, v = a + b v^2$. The contributions to the ``$a$'' term from all the
diagrams (see Eq.~\ref{eq:sigma_NLOv} for the definition) 
are listed in Table~\ref{tab:NLOa}. Both ${\cal{O}}(T^2)$
and ${\cal{O}}(T^4)$ terms contribute. (The ${\cal{O}}(v^2)$ terms can
be found from the expressions given online in Ref.~\cite{online}.) In the
non-relativistic limit, the denominator in the full expressions reduces
to the form visible in Table~\ref{tab:NLOa}:
\begin{align}
\frac{1}{(4 H^4 - 4 H^2 m_\Phi^2 + m_\Phi^4 - 4 P^2 P'^2)^n}
 & \approx
\frac{1}{(m_\chi^2 - m_f^2 + m_\phi^2)^{n}} \left[ 1 - \frac{4 n P^2
m_\phi^2} {(m_\chi^2 - m_f^2 + m_\phi^2)} \right]~, 
\nonumber \\
 & \overset{v\to 0}{\longrightarrow} 
 \qquad \frac{1}{(m_\chi^2 - m_f^2 + m_\phi^2)^{n}} \equiv \frac{1}{D^n}~,
\label{eq:D}
\end{align}
as was encountered in the LO expression for the cross section in
Eq.~\ref{eq:sigma_LOv}. It can be seen from Table~\ref{tab:NLOa} that,
in the non-relativistic limit, the ``$a$'' terms in the cross section
are proportional to the square of the fermion mass, both at order
${\cal{O}}(T^2)$ and ${\cal{O}}(T^4)$, just as in the LO case. Again,
as with the NLO calculation in the heavy-scalar limit, the potential
leading contribution at order $1/m_\phi^4$ is suppressed by a factor of
$P^2$ and contributes\footnote{Expressions for the ``$b$'' term in the cross
section are given online in Ref.~\cite{online}; note that these are not
necessarily proportional to $m_f^2$.} at the order ${\cal{O}}(P^2
T^4 m_\phi^6/(m_\chi^2 D^5)) \sim {\cal{O}}(P^2 T^4 /(m_\chi^2
m_\phi^4))$, as before. Hence, the leading contribution is again at order
${\cal{O}}(1/m_\phi^6)$, as mentioned earlier. Note that only Diagrams
2 and 4 have ${\cal{O}}(T^4)$ contributions; that from Diagram 4 is
highly suppressed due to the $1/D^5 \sim 1/m_\phi^{10}$ denominator
factor. In general, it can be seen that the leading contributions are
of order ${\cal{O}}(m_f^2 m_\chi^2 T^2/m_\phi^6)$ in the
non-relativistic limit.

\begin{table}[htp]
\centering
\setstretch{2.0}
\begin{tabular}{|c|c|c|c|} \hline
Diagram & $\gamma/f$ & $Int_{NLO}^a$ $(T^2$ contribution)
& $Int_{NLO}^a$ ($T^4$ contribution) \\ \hline
1 & $\gamma$ & $-{8 m_\chi^2 m_f^2 (m_f^2 - m_\phi^2)}/D^4$ & 0 \\
 & $f$ & ${4 m_\chi^2 m_f^2 (5 m_\chi^2 - 5 m_f^2 + m_\phi^2)}/D^4$ & 0 \\
 & Total$_{\gamma+f}$
 & ${2 m_\chi^2 m_f^2 (5 m_\chi^2 - 9 m_f^2 + 5 m_\phi^2)}/D^4$
 & 0 \\ \hline

2 & $\gamma$ & $-{8 m_\chi^2 m_f^2}/D^3$ & 0 \\
 & $f$ & $-{6 m_f^2 (2 m_\chi^2 - m_f^2)}/D^3$
 & $-\frac{21 \pi^2 T^2}{10 m_\chi^2 D^3 } {m_f^2 (2 m_\chi^2 - m_f^2)}$ \\
 & Total$_{\gamma+f}$ & $-{m_f^2 (14 m_\chi^2 - 3 m_f^2)}/D^3$ 
 & $-\frac{21\pi^2 T^2}{10 m_\chi^2 D^3 }{m_f^2(2 m_\chi^2-m_f^2)}$ \\ \hline

3 & $\gamma$ & $-{8 m_\chi^2 m_f^2 (m_f^2 - m_\phi^2)}/D^4$ & 0 \\
 & $f$ & ${4 m_\chi^2 m_f^2 (3 m_\chi^2 - 2 m_f^2 + m_\phi^2)}/D^4$ & 0 \\ 
 & Total$_{\gamma+f}$
 & ${2 m_\chi^2 m_f^2 (3 m_\chi^2 - 6 m_f^2 + 5 m_\phi^2)}/D^4$ & 0 \\ \hline

4 & $\gamma$ & ${32 m_\chi^4 m_f^2}/D^4$
 & $-\frac{56 \pi^2 T^2}{15 D^5 } m_\chi^2 m_f^2 (m_\chi^2-m_f^2)$ \\ \hline

5 & $\gamma$ & $-{16 m_\chi^2 m_f^2}/D^3$ & 0 \\ \hline

All & Total$_{\gamma+f}$
 & $\frac{1}{D^3} m_f^2 (2 m_\chi^2 + 3 m_f^2) +$
 & $-\frac{21 \pi^2 T^2}{10 m_\chi^2 D^3 } m_f^2 (2 m_\chi^2 - m_f^2)$ +\\
 & & $\frac{2}{D^4} m_f^2 m_\chi^2 (10 m_\phi^2 + 24 m_\chi^2 - 15 m_f^2)$
 & $-\frac{56 \pi^2 T^2}{15 D^5 } m_\chi^2 m_f^2 (m_\chi^2-m_f^2)$ \\ \hline
\end{tabular}
\setstretch{1.0}
\caption{\small \em The $v\to 0$ contributions from various diagrams to
the NLO cross section (the so-called ``$a$'' terms in the non-relativistic
cross section); see Eqs.~\ref{eq:sigma_NLOv} and \ref{eq:D} for
definitions. The second column lists the ${\cal{O}}(T^2)$ contributions
while the third one lists the ${\cal{O}}(T^4)$ contributions; note that
an overall factor of $T^2$ has been removed from the terms. Here $D$
is defined as $D= (m_\chi^2 - m_f^2 + m_\phi^2)$.}
\label{tab:NLOa}
\end{table}

\noindent Considering only the corresponding ``$a$'' ($v^0$) terms in the
non-relativistic limit, the relative size of the NLO contribution {\em
for each} flavour of fermion pair is given by,
\begin{align}
\frac{\sigma_{NLO}^a} {\sigma_{LO}^a} & = \frac{\pi \alpha T^2}{6
m_\phi^2} \frac{m_f^2 (22 m_\chi^2 + 3 m_f^2)}{m_f^2 m_\chi^2 }~,
\nonumber \\
 & \approx \frac{11 \pi \alpha}{3}
\frac{T^2}{m_\phi^2}~,
\label{eq:ratio}
\end{align}
where we have used $D \sim m_\phi^2$ and dropped terms in
$\sigma^{NLO}_a$ having higher powers of $D=m_\phi^2$ in the denominator.

\section{Discussions and Conclusions}

We have computed the next-to-leading order thermal contributions to the
cross section for dark matter annihilation into fermion pairs. We have
used a Lagrangian which couples the dark matter Majorana fermions
to standard model fermions via charged scalar doublets. We use the real
time formulation of thermal field theory to compute the higher order
contributions to the cross section for the annihilation of dark
matter particles\footnote{Since the dark matter particles are taken
to be Majorana fermions, the annihilation occurs via the $t$-, $u$-
and cross $tu$-channel processes; Dirac dark matter particles will have
the same interaction where the $tu$ cross terms vanish.} via $\chi \chi
\to f \overline{f}$ in a heat bath of fermions, scalars and photons at
temperature $T$. Note that the regime of interest is during the evolution
of the Universe when electroweak symmetry breaking is complete so that
we consider higher order corrections only from photons, since the $W$
and $Z$ are by now massive. We assume in addition that the scalars are
heavy, $m_\phi \gtrsim m_\chi$.

We have used the generalised Grammer and Yennie
\cite{Grammer:1973db,Yennie:1961ad, Indumathi:1996ec} technique where
the cancellation of infra-red soft divergences is straightforward. This
is achieved by separating the photon propagator into the sum of $K$
and $G$ parts. This can be realised in the thermal case as well,
where the propagators are now sums of temperature-independent and
temperature-dependent $2 \times 2$ matrices; we label the latter as
the {\sf thermal} part of the propagator since it carries an explicit
temperature dependence. It was shown earlier \cite{Sen:2020oix,
Sen:2018ybx} that the infrared (IR) divergences (which are much more
acute than at zero temperature, being linear rather than logarithmically
divergent in the soft limit) are completely contained in the $K$ photon
contribution and the $G$ photon term is finite. The infrared divergences
cancel against similar terms from real soft photon contributions; note
that both absorption from, and emission into the heat bath is required
to be included for these cancellations to occur.

While the use of the Grammer and Yennie technique (and its modification for
thermal field theories) is well-known for demonstration of the
cancellations of IR divergences in various contexts, it has been rarely
used, so far as the authors are aware, to actually compute the IR-finite
remainder. In the present work, we have computed the NLO finite
$G$ photon contributions to the dark matter annihilation cross section
$\chi \chi \to f \overline{f}$ using this technique and found its
application to be straightforward. A further simplification occurred since
kinematically it turns out that the total contribution to the NLO cross
section is actually a sum of terms where {\em any one} of the (photon,
fermion, anti-fermion) propagators can be {\sf thermal} at a time. Hence
the Grammer and Yennie approach allows a great simplification of the
problem and consequent ease of computation.

The calculation was first done by replacing the full scalar propagator
$1/(l^2 - m_\phi^2)$ by just the mass term, $1/(-m_\phi^2)$, valid when
the scalar is very heavy. This enabled us to have simpler expressions so
that the pattern of dependence of the NLO cross section on both the
temperature $T$ as well as the energy scales, $H = \sqrt{s}/2$ and $ m_f$,
was visible. The five diagrams that contribute at NLO are shown in
Fig.~\ref{fig:nlo} for the $t$-channel processes with analogous diagrams
for the $u$-channel ones.

The potential leading contribution\footnote{This is the order
at which the LO cross section contributes; see Eqs.~\ref{eq:LO},
\ref{eq:sigma_LO} and \ref{eq:sigma_LOv}.} at order $1/m_\phi^4$ in
the scalar mass (from Diagram 2) is of order ${\cal{O}}(p^2T^4/(H^2
m_\phi^4))$, which is small in the non-relativistic limit when $p^2$ is
small. Hence the leading contribution is at order $1/m_\phi^6$. It was
found that the ${\cal{O}}(T^2)$ contribution to the NLO cross section
at this order is proportional to either $(m_f^2 M^2 T^2/m_\phi^6)$ or
to $(p^2 M^2 T^2/m_\phi^6)$ where $M^2 \sim H^2$ or $m_\chi^2$. Hence
in the non-relativistic regime (at around $m_\chi/T \sim 20$) where
freeze-out occurs, the momentum-dependent term can be ignored and the
NLO cross section is proportional to the square of the fermion mass;
this dependence is also seen in the LO cross section and occurs due to
the Majorana nature of the dark matter particles and consequent helicity
suppression. It is likely that the NLO terms retain this feature due to
the $2 \to 2$ planar nature of the contributing diagrams.

The calculation was repeated, retaining the full dynamical scalar
propagator. The essential features of the calculation remain the same.
Since the expressions are very long, we present here only the leading
``$a$'' term of the cross section in the non-relativistic limit
(contribution to $\sigma v$ when $v\to 0$) in Table \ref{tab:NLOa}; the
entire solution in the relativistic case as well as their non-relativistic
approximations are listed in the Mathematica Notebooks available
online \cite{online}. We find that both the ${\cal{O}}(T^2)$ and the
${\cal{O}}(T^4)$ contributions to the ``$a$'' term are proportional
to $m_f^2$, as can be seen from Table \ref{tab:NLOa}. As with the
heavy-scalar approximation, the potential leading contribution at order
$1/m_\phi^4$ is suppressed by a factor of $p^2$ and contributes at the
order ${\cal{O}}(p^2 T^4 m_\phi^6/(m_\chi^2 D^5))$ which reduces, in
the limit $m_\phi > m_\chi$, to the order ${\cal{O}}(p^2 T^4 /(m_\chi^2
m_\phi^4))$, as before. Hence, the leading contribution is again at order
$1/m_\phi^6$, leading to a relative NLO thermal contribution which is
$\sim 10\, (\alpha T^2/m_\phi^2)$ times the LO cross section.

In summary, the NLO thermal correction terms have a $T^2$ dependence,
but are suppressed by additional powers of the (heavy scalar) propagator
as well as by the square of the fermion masses. This was discussed
in Ref.~\cite{Beneke:2014gla} where the NLO thermal cross section was
initially computed to be of order ${\cal{O}}(T^2)$, and later it was
shown that the $T^2$ contributions vanish and the leading contribution
is in fact of order ${\cal{O}}(T^4)$. This result was strengthened
in Ref.~\cite{Beneke:2016ghp} where an operator product expansion
(OPE) approach was used to show the same result (for Dirac-type
dark matter particles); here, the scalar propagator was
again reduced to just the mass term ($1/(-m_\phi^2)$), and it was
argued that ${\cal{O}}(m_f^2 T^2/m_\chi^4)$ are small compared to
${\cal{O}}(T^4/m_\chi^4)$ and were thus ignored.

In this work, we have retained the fermion masses and found that, in
the non-relativistic limit, the leading terms as $v \to 0$ at both order
${\cal{O}}(T^2)$ and ${\cal{O}}(T^4)$ are suppressed by $m_f^2$, just as
the LO cross section. The general results when $T$ may be large, so $x =
m_\chi/T \sim {\cal{O}}(1)$, are given without making the non-relativistic
approximation, in Mathematica notebooks \cite{online}. These corrections
may be significant in a freeze-in scenario in the early Universe when $T$
is larger for light dark matter candidates where $m_\chi/T \gtrsim 1$.

Note that there are two {\em independent} contributions to the
cross section in Eq.~\ref{eq:sigmavT} below, that occurs in the
Boltzmann equation for the relic densities: one from virtual photon
corrections (as has been computed here) and the other from real photon
emission/absorption processes, {\em viz.}, $\chi \chi \to f \overline{f}
(\gamma)$. As discussed in Section~\ref{sec:GY}, the soft infra-red
$\widetilde{K}$-photon contribution from the latter cancels the
corresponding divergences in the $K$-photon contribution from virtual
photon corrections order-by-order in the theory so that the finite cross
section for $\chi \chi \to f \overline{f}$ could be calculated simply
by calculating the $G$-photon virtual contribution.

The finite part of the real photon emission/absorption processes
(obtained by computing the $\widetilde{G}$ contribution to $\chi \chi \to
f \overline{f} (\gamma)$, analogous to the $\widetilde{K}$ contribution
discussed in Eqs.~\ref{eq:KGtilde}, \ref{eq:realphspace} and below) also
contributes to the cross section that appears in the collision term of
the Boltzmann equation. This has not been computed here, although we have
used the result from Ref.~\cite{Beneke:2016ghp} where the cancellation
of the collinear terms between the virtual and real photon contributions
has been shown; see, for instance, discussion in Eqs.~\ref{eq:sigma_1g},
\ref{eq:Intt1g}, and below. The total cross section for dark matter
annihilation is then the sum of these two contributions. Thermal
corrections to such cross sections can become important in the early
Universe where the thermally averaged cross section, $\langle \sigma v
\rangle$, appears as the collision term in the Boltzmann equation for
the evolution of the dark matter phase space densities, or equivalently,
their yields. We have
\begin{align}
\langle \sigma \cdot v_{\hbox{M\o l}} \rangle \equiv
\frac{\int \sigma \cdot v_{\hbox{M\o l}} \exp[-E_1/T] \exp[-E_2/T] {\rm d}^3 p_1
{\rm d}^3 p_2}
{\int \exp[-E_1/T] \exp[-E_2/T] {\rm d}^3 p_1 {\rm d}^3 p_2}~.
\end{align}
Here $p_1, p_2$ are the momenta of the annihilating dark matter
particles, and $v_{\hbox{M\o l}} = \sqrt{(p_1 \cdot p_2)^2 - m_\chi^2}$.
The Boltzmann approximation for the dark matter number densities has been
used. For cross sections of the form $\sigma v = a + b v^2$, we have
\begin{align}
\langle \sigma v \rangle = a + \frac{3}{2} b T~.
\label{eq:sigmav}
\end{align}
Hence, the thermal corrections that we have computed at NLO will yield a
correction,
\begin{align}
\langle \sigma_{NLO} v \rangle & = \langle (a_1 T^2 + a_2 T^4) + v^2 (b_1 T^2
+ b_2 T^4) \rangle~, \nonumber \\
 & = (a_1 T^2 + a_2 T^4) + \frac{3}{2} (b_1 T^3 + b_2 T^5)~.
\label{eq:sigmavT}
\end{align}
The corrections from the NLO thermal contribution to ``$b$'' (the terms
proportional to $v^2$ and higher powers) can therefore be neglected,
although $p$-wave enhancement can occcur \cite{Beneke:2024iev,
Qiu:2024iyo} due to the Sommerfeld effect. These thermal corrections
to the cross section $\sigma(s)$ can alter the yield equation and
hence the final relic density, especially in freeze-in scenarios. This
is acquiring importance \cite{Becker:2023vwd, Biondini:2020ric} in
view of the increasingly precise measurements of the relic density
\cite{Planck:2018vyg, Amendola:2016saw}. It will be interesting to
pursue this further, within the frame-work of the Grammer and Yennie
formulation where the calculation is simpler; this is beyond the scope
of the present work.

\appendix
\renewcommand{\theequation}{\thesection.\arabic{equation}}
\setcounter{equation}{0}

\section{Feynman rules in thermal field theory}
\label{sec:feyn}

The scalar propagator is given by
\begin{equation}
i {{S}}_{\rm scalar}^{t_a, t_b} (p,m) = \left(
	\begin{array}{cc}
	\Delta(p) & 0 \\
	0 & \Delta^*(p) \end{array} \right) +
	2 \pi \delta(p^2-m^2) n_{\rm B}(\vert p^0 \vert) 
	\left(\begin{array}{cc}
	1 & e^{\vert p^0 \vert /(2T)} \\
	e^{\vert p^0 \vert /(2T)} & 1 
	\end{array} \right)~,
\label{eq:sprop}
\end{equation}
where $\Delta(p) = i/(p^2-m^2+i\epsilon)$, and $t_a, t_b~(=1,2)$ refer
to the field's thermal type. We shall refer to the first term as the
{\sf temperature-independent} part and the second as the {\sf thermal}
part in the main text since it carries explicit temperature dependence;
note that the latter contributes only on mass-shell. Only the {\sf
thermal} parts can convert type-1 to type-2 fields, and vice versa;
type-1 fields are the physical fields.

In the Feynman gauge, the photon propagator corresponding to
a momentum $k$ is given by
\begin{align}
i {\cal D}^{t_a,t_b}_{\mu\nu} (k) =
-g_{\mu\nu} i {D}^{t_a,t_b} (k) =
- g_{\mu\nu} \, i {{S}}_{\rm scalar}^{t_a, t_b} (k,0)~,
\label{eq:thermalD}
\end{align}
and the fermion propagator at zero chemical potential is given by
\begin{align} \nonumber
i {\cal{S}}_{\rm fermion}^{t_a, t_b} (p,m) & = \left(
	\begin{array}{cc}
	S & 0 \\
	0 & S^*
	\end{array} \right)
	- 2 \pi S' \delta(p^2-m^2) n_{\rm F}(\vert p^0 \vert) 
	\left(\begin{array}{cc}
	1 & \epsilon(p_0) e^{\vert p^0 \vert /(2T)} \\
	-\epsilon(p_0) e^{\vert p^0 \vert /(2T)} & 1 
	\end{array} \right)~, \\
 & \equiv (\slashed{p}+m) \left(\begin{array}{cc}
 	F_p^{-1} & G_p^{-1} \\
        -G_p^{-1} & F_p^{*-1}
	\end{array} \right)~, \nonumber \\
 & \equiv (\slashed{p}+m) \overline{S}^{t_a,t_b}(p,m)~,
\label{eq:fprop}
\end{align}
where $S = i/(\slashed{p} -m+i\epsilon)$, and $S' = (\slashed{p} +m)$;
hence the entire fermion propagator is proportional to $(\slashed{p}+m)$,
just as at $T=0$. It can be seen that all propagators are a sum of {\sf
temperature-independent} and {\sf thermal} parts. The fermionic
number operator,
\begin{equation} 
n_{\rm F}(\vert {p^0} \vert) \equiv
\frac{1}{\exp\{\vert p^0 \vert/T\} + 1} \;
\stackrel{p^0 \to \,0}{\longrightarrow} \; \frac{1}{2}~,
\label{eq:nF}
\end{equation}
is well-defined in the soft limit; however, the bosonic number operator
contributes an additional power of ${k^0}$ in the denominator to the
photon propagator in the soft limit:
\begin{equation} 
n_{\rm B}(\vert k^0 \vert) \equiv 
	\frac{1}{\exp\{\vert k^0 \vert/T\} - 1} 
	\; \stackrel{k^0 \to \,0}{\longrightarrow} \;
	\frac{T}{\vert k^0 \vert}~,
\label{eq:nB}
\end{equation}
so that the leading IR divergence in the finite temperature part is
linear rather than logarithmic as was the case at zero temperature. (The
cancellation of IR divergence thus involves cancellation of the leading
linear divergence as well as the logarithmic sub-divergence.)

\begin{figure}[htp]
\centering
\includegraphics[width=\textwidth]{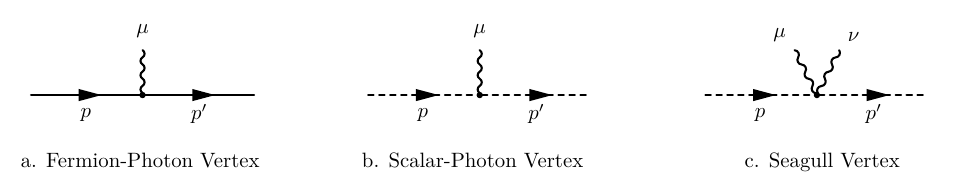}
\caption{\small \em Allowed vertices for fermion--photon and
scalar--photon interactions.}
\label{fig:feynman}
\end{figure}

The fermion--photon vertex factor is given by
$(-ie\gamma_\mu)(-1)^{t_\mu+1}$, where $t_\mu=1,2$ for the
type-1 and type-2 vertices.
The scalar--photon vertex factor is $[-ie(p_\mu
+ p'_\mu)](-1)^{t_\mu+1}$ where $p_\mu$ ($p'_\mu$) is the 4-momentum of
the scalar entering (leaving) the vertex, while the 2-scalar--2-photon
{\em seagull} vertex factor (see Fig.~\ref{fig:feynman}) is
$[+2ie^2 g_{\mu\nu}](-1)^{t_\mu+1}$ (the factor `2' is dropped for a
{\em tadpole} vertex). At a given vertex, all fields are of the same
type.

The bino-scalar-fermion vertex factor is $i \lambda P_L$; for details
on Feynman rules for Majorana particles at zero temperature, see
Ref.~\cite{Denner:1992vza}. An overall negative sign applies as usual to
the type-2 bino vertex; again all fields at a vertex are of the same type.

\subsection{Some identities at finite temperature}
\label{app:fidentities}

Various identities useful for fermions are given in Ref.~\cite{Sen:2020oix}
and are reproduced here for completeness. Note that
Eqs.~\ref{eq:sprop}, \ref{eq:thermalD} and \ref{eq:fprop} lead to
\begin{align}
(\slashed{p}-m) ~ i{S}_{p}^{t_a,t_b} & = i (-1)^{t_a+1}
\delta_{t_a,t_b}~,
\label{eq:propid}
\end{align}
where we have used the compressed notation,
$i{S}^{t_a,t_b}(p,m) \equiv i S^{t_a,t_b}_p$.

Consider the insertion of the $\mu$ vertex of the additional virtual
photon with momentum $k$ between vertices $\mu_{q+1}$
and $\mu_q$ on the $p'$ fermion leg; see Fig.~\ref{fig:finite}.
The momentum of the photon at the
vertex $\mu_q$ is $l_q$, with
Lorentz index $\mu_q$, and thermal
type-index $t_q$. Hence the momentum of the fermion leg to the left
of the vertex $\mu_q$ is
$p'+\sum_{i=1}^{q}l_i$ which we denote as
$p'+\sum_q$. Using Eq.~\ref{eq:propid}, we have,
\begin{align}
S\strut^{t_q,t_\mu}_{p'+\sum\limits_q} \, \slashed{k} \,
	S\strut^{t_\mu,t_{q+1}}_{p'+\sum\limits_q+k} & =
	(-1)^{t_\mu+1} \left[S\strut^{t_q,t_{q+1}}_{p'+\sum\limits_q}
	\delta_{t_\mu,t_q+1} - S\strut^{t_q,t_{q+1}}_{p'+\sum\limits_q+k}
	\delta_{t_\mu,t_q} \right]~.
\label{eq:fid}
\end{align}
If the photon vertex is inserted to the right of the vertex labelled
`1' on the fermion leg with momentum $p'$, we have,
\begin{align}
\overline{u}(p') \slashed{k} S\strut^{t_\mu,t_1}_{p'+k} & =
\overline{u}(p') (-1)^{t_\mu+1} \delta_{t_\mu,t_1}~,
\label{eq:endf}
\end{align}
since $\slashed{p}'u(p') = m u(p')$. Similar relations hold for the
insertion of of the virtual $K$ photon at a vertex $\nu$ on the (anti-)fermion
$p$ leg since $\slashed{p}\, u(p) = m\, u(p)$ as well.

\section{Matrix elements for $\chi \chi \to f \overline{f}$ at NLO}
\label{sec:appb}

As described in the text, the matrix elements are sums of the
contributions when any one of the photon, fermion or anti-fermion
propagators is {\sf thermal}, that is, contributes via its explicitly
temperature-dependent part. The $t$- and $u$- channel matrix elements
for Diagram 1 when either the photon or fermion propagator is {\sf
thermal} is given in the text, in Eqs.~\ref{eq:fig1pt}, \ref{eq:fig1pu},
\ref{eq:fig1ft}, and \ref{eq:fig1fu}. The remaining matrix elements are
listed below. In the equations that follow, the indices
$(\hbox{Diagram}\,i, \gamma)$,
$(\hbox{Diagram}\,i, f)$ and
$(\hbox{Diagram}\,i, \overline{f})$ refer to the contribution from
Diagram $i$ ($i = 1$--5), when the photon, fermion, or anti-fermion
respectively contribute via the explicitly {\sf thermal} parts of
their propagators.

The {\sf thermal} photon contributions from the last four diagrams
in Fig.~\ref{fig:nlo} are given by
\begin{align}
{\cal{M}}^{t}_{NLO} (\hbox{\small Diagram}\, 2, \gamma)
& = 
\int \frac{\hbox{d}^4k}{(2\pi)^4}
\frac{i e^2 \lambda ^2}{4 k \cdot p' k\cdot p}
\left(2\pi \delta(k^2) n_B(\vert k^0 \vert) \right) (i \Delta(l+k)) \nonumber \\
 & \qquad \left[\left(\overline{u}({p'},{m_f})\gamma_\mu \ 
\left(\slashed{k}+\slashed{p'}+m_f\right) P_R\, u({q'},{m_\chi}) \right)
\right. \nonumber \\
 & \qquad \left. \left(\overline{v}(q,{m_\chi}) P_L
\left(\slashed{k}-\slashed{p}+m_f\right) \gamma_\nu \, v(p,{m_f}) \right)
\right] G_k^{\mu\nu}(p',p)~, \nonumber \\
{\cal{M}}^{t}_{NLO} (\hbox{\small Diagram}\, 3, \gamma) & = 
\int \frac{\hbox{d}^4k}{(2\pi)^4}
\frac{i e^2 \lambda ^2}{2 k\cdot p}
\left(2\pi \delta(k^2) n_B(\vert k^0 \vert) \right) \Delta(l) \Delta(l+k) \nonumber \\
 & \qquad \left[\left(\overline{u}({p'},{m_f}) P_R\, u({q'},{m_\chi}) \right)
 (k - 2p + 2 q)_\mu \right. \nonumber \\
  & \left. \qquad \left(\overline{v}(q,{m_\chi}) P_L
\left(\slashed{k}-\slashed{p}+m_f\right) \gamma_\nu \, v(p,{m_f}) \right)
\right] G_k^{\mu\nu}(p,p)~, \nonumber \\
{\cal{M}}^{t}_{NLO} (\hbox{\small Diagram}\, 4, \gamma) & =
\int \frac{\hbox{d}^4k}{(2\pi)^4} (i e^2 \lambda ^2)
\left(2\pi \delta(k^2) n_B(\vert k^0 \vert) \right) \Delta(l) i \Delta(l+k) \Delta(l)
\nonumber \\
 & \qquad \left[\left(\overline{u}({p'},{m_f}) P_R\, u({q'},{m_\chi}) \right)
 (k - 2p + 2 q)_\mu \, (k - 2p + 2 q)_\nu \right. \nonumber \\
& \qquad \left. \left(\overline{v}(q,{m_\chi}) P_L\, v(p,{m_f}) \right)
\right] G_k^{\mu\nu}(p,p)~, \nonumber \\
{\cal{M}}^{t}_{NLO} (\hbox{\small Diagram}\, 5, \gamma) & = 
\int \frac{\hbox{d}^4k}{(2\pi)^4}
(i e^2 \lambda ^2)
\left(2\pi \delta(k^2) n_B(\vert k^0 \vert) \right) \Delta(l) \Delta(l) \nonumber \\
 & \qquad \left[\left(\overline{u}({p'},{m_f}) P_R\, u({q'},{m_\chi}) \right)
 \left(\overline{v}(q,{m_\chi}) P_L\, v(p,{m_f}) \right)
\right] G_k^{\mu\nu}(p,p)~,
\end{align}
for the $t$-channel, and appropriately crossed ones for the $u$-channel,
for example, as shown in Eq.~\ref{eq:fig1pu} for the contribution from
the first diagram. The matrix elements for the case when the fermion or
anti-fermion is {\sf thermal} arise only from Diagrams 1, 2, and 3. They
are given by
\begin{align}
{\cal{M}}^{t}_{NLO} (\hbox{\small Diagram}\, 2, f) & =
\int \frac{\hbox{d}^4t} {(2\pi)^4}
\frac{i e^2 \lambda ^2
\,g^{\mu\nu}}{4 (m_f^2-p' \cdot t)\,(p\cdot p'+m_f^2 - t \cdot
(p+p'))} \nonumber \\
 & \qquad \left(-2\pi \delta(t^2-m_f^2)n_F(\vert t^0\vert) \right)
i \Delta(l+k) \left[\left(\overline{u}({p'},{m_f})\gamma_\mu
\left(\slashed{t}+m_f\right) \right. \right. \nonumber \\
 & \left. \left. \qquad P_R\, u({q'},{m_\chi}) \right)
\left(\overline{v}(q,{m_\chi}) P_L \,
\left(\slashed{t}-\slashed{p'} - \slashed{p}\right) \gamma_{\nu }
v(p,{m_f}) \right) \right]~,
\end{align}
\begin{align}
{\cal{M}}^{t}_{NLO} (\hbox{\small Diagram}\, 2, \overline{f}) & =
\int \frac{\hbox{d}^4t} {(2\pi)^4}
\frac{i e^2 \lambda ^2
\,g^{\mu\nu}}{4(m_f^2-p \cdot t)\,(p\cdot p'+m_f^2 - t \cdot
(p+p'))} \nonumber \\
 & \qquad \left(-2\pi \delta(t^2-m_f^2)n_F(\vert t^0\vert) \right)
 i \Delta(l+k) \left[\left(\overline{u}({p'},{m_f})\gamma_\mu
\left(-\slashed{t}+\slashed{p'} + \slashed{p}+m_f\right)
\right. \right. \nonumber \\
 & \left. \left. \qquad P_R\, u({q'},{m_\chi}) \right)
 \left(\overline{v}(q,{m_\chi}) P_L \,
\left(-\slashed{t} + m_f \right) \gamma_{\nu }
v(p,{m_f}) \right) \right]~, \\
{\cal{M}}^{t}_{NLO} (\hbox{\small Diagram}\, 3, \overline{f}) & =
\int \frac{\hbox{d}^4t} {(2\pi)^4}
\frac{i e^2 \lambda ^2
\,g^{\mu\nu}}{2(m_f^2 - t \cdot p)}
 \left(-2\pi \delta(t^2-m_f^2)n_F(\vert t^0\vert) \right) \Delta(l)
 \Delta(l+k) \nonumber \\
 & \qquad \left[\left(\overline{u}({p'},{m_f})
P_R\, u({q'},{m_\chi}) \right)
 \left(-t-p+2 q\right)_{\mu } \right. \nonumber \\
 & \qquad \left. \left(\overline{v}(q,{m_\chi}) P_L \,
 \left(-\slashed{t}+m_f \right) v(p,{m_f}) \right) \right]~.
\end{align}
Here $f$ corresponds to the fermion being thermal in Diagram 2 and
$\overline{f}$ corresponds to the anti-fermion being thermal in Diagrams
2 and 3 respectively. We use the same replacement technique here as
explained in Eqs.~\ref{eq:tsub} and \ref{eq:Ot} for the {\sf thermal}
fermion contribution to Diagram 1. For the matrix element of Diagram 2,
we have substituted $t = p'+k$ as for Diagram 1; and we have substituted
$t=p-k$ when the anti-fermion is thermal. Analogous expressions hold
for the $u$-channel matrix elements.

\newpage
\printbibliography

\end{document}